\documentclass[a4paper,11pt]{article}
\usepackage{jheppub} % for details on the use of the package, please see the JINST-author-manual
\usepackage{lineno}
\usepackage{mathabx}
\usepackage[compat=1.1.0]{tikz-feynman}
\usepackage{slashed}
\usepackage[normalem]{ulem}
\usepackage{subcaption}

\arxivnumber{2410.xxxx} % if you have one

\preprint{\begin{flushright}
    UT-WI-41-2024\\
\end{flushright}}

\title{Searching for Hidden Sector Particles at Neutrino Telescopes}

\author[a]{Sagar Airen}
\author[a]{Zackaria Chacko}
\author[b]{Can Kilic}
\author[b]{Ram Purandhar Reddy Sudha}
\affiliation[a]{Maryland Center for Fundamental Physics, Department of Physics, University of Maryland,
     College Park, MD 20742, USA}
\affiliation[b]{Theory Group, The Weinberg Institute for Theoretical Physics,
University of Texas at Austin, Austin, TX 78712, USA}

\emailAdd{sairen@umd.edu}
\emailAdd{zchacko@umd.edu}
\emailAdd{kilic@physics.utexas.edu}
\emailAdd{ramreddy@utexas.edu}

\abstract{ 
We explore the possibility of directly detecting light, long-lived hidden sector particles at the IceCube neutrino telescope. Such particles frequently arise in non-minimal hidden sectors that couple to the Standard Model through portal operators. We consider two distinct scenarios. In the first scenario, which arises from a neutrino portal interaction, a hidden sector particle is produced inside the detector by the collision of an energetic neutrino with a nucleon, giving rise to a visible cascade. This new state then decays into a hidden sector daughter, which can naturally be long-lived. The eventual decay of the daughter particle back to Standard Model states gives rise to a second cascade inside the detector. This scenario therefore gives rise to a characteristic ``double bang" signal arising from the two distinct cascades. In the second scenario, which arises from a hypercharge portal interaction, a hidden sector particle is produced outside the detector by the collision of an atmospheric muon with a nucleon. This new state promptly decays into a pair of hidden sector daughters that are long-lived. If both daughters decay into Standard Model states inside the detector, we again obtain a double-bang signal from the two distinct cascades. We explore the reach of IceCube for these two scenarios and show that it has the potential to significantly improve the sensitivity to hidden sector models in the mass range from about a GeV to about 20 GeV.
}

\begin{document}
\maketitle
\flushbottom

\section{Introduction}
\label{sec:intro}

The existence of a hidden sector composed of particles that do not transform under any of the Standard Model (SM) gauge groups represents a highly intriguing possibility for new physics. Such hidden sectors can offer solutions to several puzzles of the SM, including the little hierarchy problem~\cite{Chacko:2005pe,Barbieri:2005ri,Craig:2015pha,Cohen:2018mgv,Cheng:2018gvu}, the origin of the neutrino masses~\cite{Minkowski:1977sc,Yanagida:1979,Gell-Mann:1979vob,Glashow:1979,Mohapatra:1979ia}, the nature of dark matter~\cite{Silveira:1985rk,McDonald:1993ex,Burgess:2000yq,Dodelson:1993je,Pospelov:2007mp,Feng:2008mu}, and the origin of the baryon asymmetry~\cite{Fukugita:1986hr,Luty:1992un,Farina:2016ndq,Feng:2020urb,Kilic:2021zqu}.
Given that the experimental constraints on particles that are neutral under the SM are relatively weak, the states in a hidden sector could have masses below the weak scale, provided that their couplings to the SM are sufficiently small. Understanding the extent to which current and future experiments will be able to probe such light hidden sectors remains an open problem.

The experimental signals associated with a hidden sector depend on how it couples to the SM, and also on its particle content. Hidden sectors interact with the SM through interactions of the form $\mathcal{O}_{\rm SM}\mathcal{O}_{\rm HS}$, which couple gauge invariant SM operators ${\mathcal O}_{\rm SM}$ to operators $\mathcal{O}_{\rm HS}$ in the hidden sector. The SM operators ${\mathcal O}_{\rm SM}$ represent ``portals'' to the hidden sector.
From the point of view of experiment, the most promising portals are the Higgs portal ${\mathcal O}_{\rm SM} \equiv H^{\dag}H$, the neutrino portal ${\mathcal O}_{\rm SM} \equiv L H$, and the hypercharge portal ${\mathcal O}_{\rm SM} \equiv B_{\mu\nu}$. This is because these portals have the lowest scaling dimensions and can therefore be probed with the greatest experimental sensitivity. 

For each of these portals, the experimental signals of the case when the hidden sector consists of just a single particle have been well studied. These include searches for scalar particles that mix with the Higgs boson, heavy neutral leptons that couple through the neutrino portal, and dark photons that mix with the hypercharge gauge boson. However, in general, the hidden sector could consist of a large number of states with sizable interactions between them. The experimental signals of this larger class of hidden sectors have received much less attention.

If the interactions between the different particles in such a larger hidden sector are sizable, once a hidden sector particle has been produced, it is likely to decay promptly down to the lightest state in that sector, the ``Lightest Hidden Sector Particle" (LHSP). Since the LHSP is forbidden by kinematics from decaying to other hidden sector states, it must either be stable or decay back to the SM. If the couplings of the LHSP to the SM are small, or if its decays are suppressed because of symmetry considerations, it can be long-lived on detector time scales, potentially giving rise to striking signals in experiments.

Because of their enormous size, neutrino telescopes offer unique opportunities for directly detecting exotic long-lived particles predicted by models of new physics. At present, the world's largest neutrino telescope is at the IceCube Neutrino Observatory, located at the South Pole. Examples of long-lived particles that IceCube is sensitive to include staus in supersymmetric theories~\cite{Albuquerque:2003mi,Albuquerque:2006am,Ahlers:2006pf,Reno:2007kz,Ando:2007ds,Meighen-Berger:2020eun}, Kaluza-Klein particles in models of extra dimensions~\cite{Albuquerque:2008zs}, magnetic monopoles~\cite{Posselt:2011gyh,ObertackePollmann:2016uvi}, dark photons~\cite{Feng:2015hja,Green:2018qwo}, and heavy singlet neutrinos that mix with the SM neutrinos~\cite{Coloma_2017,Vannerom:2022cpf}. Dedicated searches have been performed at IceCube for staus and Kaluza-Klein particles~\cite{Kopper:2015rrp,IceCube:2021kod}, magnetic monopoles~\cite{IceCube:2012khj,IceCube:2014xnp,IceCube:2015agw,IceCube:2021eye}, and heavy singlet neutrinos~\cite{Fischer:2022zwu,Fischer:2024vmw}. 

In this paper, we explore the possibility of directly detecting light hidden sector particles at IceCube. We consider two distinct scenarios. In the first scenario, which arises from neutrino portal interactions, a hidden sector particle is produced by the collision of an energetic neutrino with a nucleon {\bf inside} the detector volume, resulting in a cascade. This new particle is then assumed to decay into a final state containing a hidden sector daughter, which can naturally be long-lived. If the eventual decay of the daughter particle back to the SM occurs inside the detector, it gives rise to a second cascade. This scenario can therefore give rise to distinctive ``double bang" events, characterized by two or more separate cascades. 

This signal has been studied earlier for the case when the hidden sector consists of just a singlet fermion coupled through the neutrino portal~\cite{Coloma_2017}. However, it was found to provide only a limited improvement in sensitivity over the existing experimental constraints. This can be traced to the fact that, in the regime where the production cross section for the fermion is sizable, its lifetime is not long enough to easily distinguish the two bangs. However, in the case when the LHSP is a scalar, its decay width into SM fermions is helicity suppressed because of angular momentum considerations, so that it can naturally be long-lived. Therefore, this scenario gives rise to well-separated bangs that are more easily distinguished.

In the second scenario, which arises from hypercharge portal interactions, a hidden sector particle is created by the collision of an atmospheric muon with a nucleon {\bf outside} the detector. This new particle is then assumed to decay into a final state that contains two long-lived hidden sector daughters.
If both daughter particles decay back to SM states inside the detector volume, they give rise to two distinct cascades, again resulting in a double bang signal. 
We focus on events in which the initial collision of the muon with the nucleus occurs above the detector because the flux of energetic muons falls rapidly with the depth below the ground.

For each of the neutrino and hypercharge portals, we explore the reach of IceCube for a simplified model of a hidden sector that gives rise to the corresponding double bang signal. We find that for the neutrino portal, IceCube has the potential to significantly improve on the current sensitivity for hidden sector particles in the mass range from a few GeV to about 20 GeV. In the case of the hypercharge portal, the corresponding mass range extends from about a GeV to about 10 GeV.

The outline of this paper is as follows. In Section~\ref{sec: Neutrino Portal}, we assess the reach of IceCube for a simplified model of a hidden sector that couples to the SM through the neutrino portal and gives rise to a neutrino-initiated double bang signal. In Section~\ref{sec: HyperCharge Portal}, we consider a simplified model of a hidden sector that couples to the SM through the hypercharge portal, giving rise to a muon-initiated double bang signal, and determine the sensitivity of IceCube to this scenario. We conclude in Section~\ref{sec: Conclusion}.

\section{The Neutrino Portal}
\label{sec: Neutrino Portal}
\subsection{Model Description}
\label{sec: NP_Model}

In this section, we consider a simplified model of a hidden sector that interacts with the SM through the neutrino portal and determine the expected event rate at IceCube. The hidden sector contains three Dirac fermions $N^{\alpha}$ that play the role of singlet neutrinos. Here $\alpha = 1,2,3$ is a hidden sector flavor index. We can decompose the singlet neutrinos $N^\alpha$ into left and right chiral components, $N^\alpha \equiv (N_L^\alpha, N_R^\alpha)$. The $N^\alpha$ are neutral under the SM gauge interactions and mediate the portal interaction,
\begin{equation}
    \mathcal{L} \supset \bar{N}i \gamma^\mu \partial_\mu N -m_N \bar{N}N -\left(\lambda \bar{L}\widetilde{H}N_R + {\rm h.c.}\right) \; ,
\label{eq: SM_HS_int}
\end{equation}
where we have suppressed flavor indices and $\widetilde{H} \equiv i\sigma_2 H^*$. In what follows, we restrict our attention to just a single flavor of SM fermions and singlet neutrinos, but we will generalize to the realistic case of three flavors later. 

In addition to the Dirac fermions $N$, we assume that the hidden sector contains a scalar $\sigma$, which is taken to be the LHSP. The hidden sector particles $N$ and $\sigma$ interact through a Yukawa coupling,
\begin{equation}
    \mathcal{L} \supset \frac{1}{2}\partial_\mu \sigma \partial^\mu \sigma - \frac{1}{2}m^2_\sigma \sigma^2 -y \sigma \bar{N}N \; .
\label{eq: sigma_N_int}
\end{equation}
After electroweak symmetry breaking, the neutrino portal interaction results in mass mixing between the Dirac fermion $N$ and the SM neutrinos.
%Due to the Majorana and Dirac mass terms for $N$, the masses of the SM %neutrinos are generated through the inverse seesaw mechanism as:
%\begin{equation}
%    m_\nu = \mu \left(\frac{\lambda v_{\rm EW}}{m_N}\right)^2
%\end{equation}
The mixing between the singlet and active neutrinos is conveniently parametrized in terms of the parameter $U_{N\ell} \equiv {\lambda v_{\rm EW}}/{m_N}$. Since constraints from precision measurements already restrict $|U_{N\ell}|^2 \lesssim 10^{-3}$,  we will restrict our analysis to the parameter range where $|U_{N\ell}|^2 \ll 1$. As a consequence of the mixing, the heavier mass eigenstate, which is mostly composed of $N$, inherits from the neutrinos couplings to the $W$ and $Z$ bosons of the SM, suppressed by a factor of $U_{N\ell}$, 
\begin{equation}
    \mathcal{L} \supset -g_Z U_{N\ell} \overline{\nu_L} \slashed{Z} N_L -  g_W U_{N\ell} \overline{\ell} \slashed{W}^{-} N_L + \mathrm{h.c.}\; .
\label{eq: IR Lagrangian_1}
\end{equation} 
In addition, through the mixing, the light mass eigenstate, which is mostly composed of the SM neutrino, obtains couplings to the scalar $\sigma$ from the Yukawa interaction in Eq.~(\ref{eq: sigma_N_int}),
\begin{equation}
    \mathcal{L} \supset - yU_{N\ell} \sigma \overline{\nu_L} N_R + \mathrm{h.c.}\; .
\label{eq: IR Lagrangian_2}
\end{equation}
At this point, we introduce a small Majorana mass term for the $N$, 
\begin{equation}
    \mathcal{L} \supset -\frac{\mu}{2}\bar{N}N^c + {\rm h.c.}\; ,
\label{eq:mu_term}    
\end{equation}
where $\mu \ll \lambda v_{\rm EW}$ and $\psi^c \equiv \mathcal{C}{\overline{\psi}}^T$ denotes the charge-conjugate of Dirac spinor $\psi$. This generates Majorana masses for the neutrinos through the inverse seesaw mechanism~\cite{Mohapatra:1986aw,Mohapatra:1986bd},
\begin{equation}
    m_\nu = \mu \left(\frac{\lambda v_{\rm EW}}{m_N}\right)^2 \;.
\end{equation}
It also gives rise to an additional coupling of the scalar $\sigma$ to the light neutrinos, 
\begin{equation}
    \mathcal{L} \supset - y U^2_{N\ell} \left(\frac{\mu}{m_N}\right) \sigma \overline{\nu_L} \nu^c_L + {\rm h.c.}\; .
\label{eq: IR Lagrangian_3}
\end{equation}

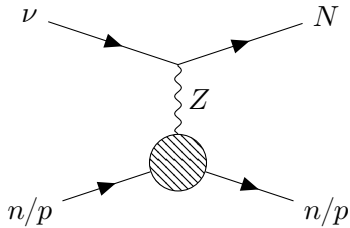
\begin{figure}[h!]
    \centering
  \begin{tikzpicture}[scale= 0.65]
   \begin{feynman}
   \vertex (a) at (-3,2){$\nu$};
  \vertex (b) at (3,2){$N$};
   \vertex (i1) at (0,1);
   \vertex[blob] (i2) at (0,-1){};
   \vertex (c) at (-3,-2){$n/p$};
   \vertex (d) at (3,-2){$n/p$};
  \diagram*{
  
  (i1)--[boson, edge label=$Z$](i2),
   (a)--[fermion](i1),
   (i1)--[fermion](b),
   (c)--[fermion](i2),
   (i2)--[fermion](d),
  };
  \end{feynman};
  
  \end{tikzpicture}
    \caption{Production of singlet neutrinos in neutrino-nucleon scattering.}
    \label{fig: Nprod}
\end{figure}

\begin{figure}[h!]
    \centering
    \includegraphics[width=0.7\linewidth]{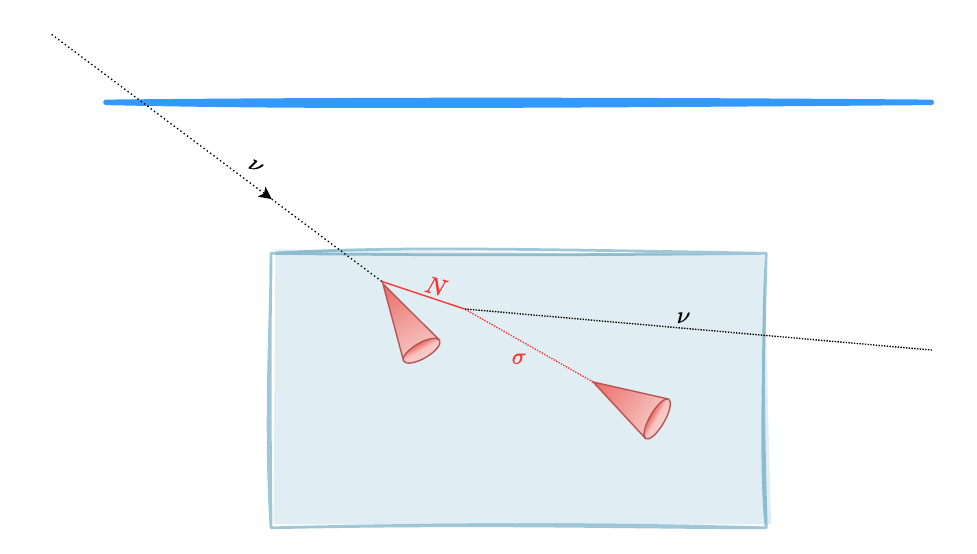}
    \caption{Illustration of the neutrino initiated double bang signal in the case of the neutrino portal model. A singlet neutrino $N$ is produced through neutrino-nucleon scattering, resulting in the first cascade. The $N$ promptly decays into $\sigma + \nu$. The long-lived $\sigma$ decays into a final state containing visible particles, resulting in the second cascade.}
    \label{fig: DB_illustration}
\end{figure}

This simple neutrino portal model can give rise to double bang signals at IceCube. From the first term in Eq.~(\ref{eq: IR Lagrangian_1}), singlet neutrinos $N$ can be produced at IceCube through deep-inelastic scattering of atmospheric neutrinos and nucleons, as shown in Fig.~\ref{fig: Nprod}. This results in energy being deposited at the site where the $N$ is produced, giving rise to a cascade. An $N$ that has been produced promptly decays into a $\sigma$ and a $\nu$. Since the scalar $\sigma$ is long-lived in most of the parameter space of interest (see Section~\ref{sec: NP_prod_decay}), it travels a significant distance in the detector before decaying into SM particles. If these decay products include hadrons, electrons, or taus, this gives rise to a distinct second cascade in the detector. This process is illustrated in Fig.~\ref{fig: DB_illustration}.

We now consider the realistic case of three generations of SM fermions and singlet neutrinos. Then, the Dirac mass term $m_N$ for the singlet neutrinos in Eq.~(\ref{eq: SM_HS_int}), the couplings $\lambda$ and $y$ 
in Eqs.~(\ref{eq: SM_HS_int}) and (\ref{eq: sigma_N_int}), and the Majorana mass parameter $\mu$ in Eq.~(\ref{eq:mu_term}), all generalize to $(3 \times 3)$ matrices in flavor space, $m_{N, \alpha \beta}$, $\lambda_{i \alpha}$, $y_{\alpha \beta}$ and $\mu_{\alpha \beta}$ respectively. Here $i=1,2,3$ represents a flavor index for the SM leptons $L^i$ and $\alpha =1,2,3$ a flavor index for the singlet neutrinos $N^\alpha$. For simplicity, we will assume that the singlet neutrinos $N^\alpha$ are approximately degenerate in mass and that the couplings $\lambda_{i \alpha}$ and $y_{\alpha \beta}$ are flavor universal, so that $m_{N, \alpha \beta} \; \propto \; \delta_{\alpha \beta}$, $\lambda_{i \alpha} \; \propto \; \delta_{i \alpha}$ and $y_{\alpha \beta} \; \propto \; \delta_{\alpha \beta}$. Then, in this framework. the nontrivial flavor structure of the mass matrix for the light neutrinos arises from the Majorana mass term for the singlet neutrinos, $\mu_{\alpha \beta}$, which is not taken to be flavor universal. With these assumptions, any lepton flavor violating process requires at least one insertion of $\mu$, and is therefore suppressed by the light neutrino masses. The model is therefore consistent with the existing constraints on lepton flavor violation.

\subsection{Production and Decay of Hidden Sector Particles}
\label{sec: NP_prod_decay}

\begin{figure}[h!]
    \centering
    \begin{subfigure}{0.45\textwidth}
        \includegraphics[width=\linewidth]{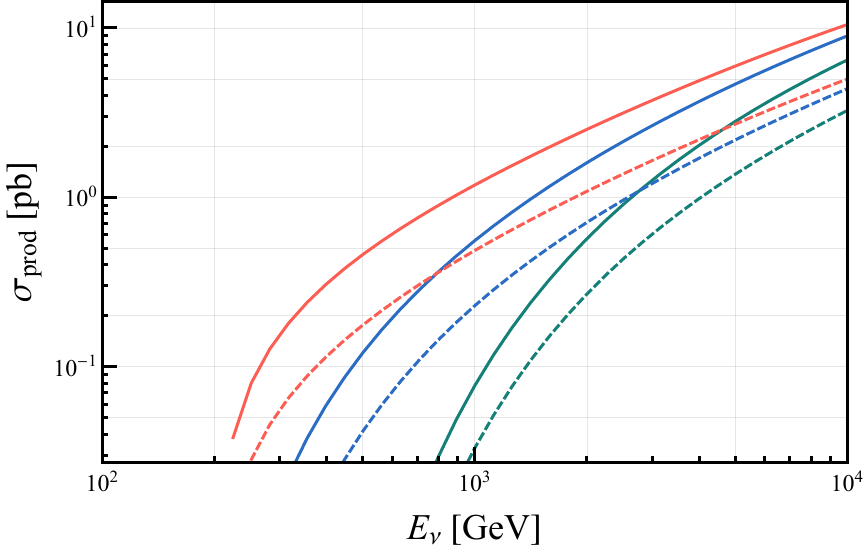}
    \end{subfigure}
    \hfill
    \begin{subfigure}{0.45\textwidth}
        \includegraphics[width=\linewidth]{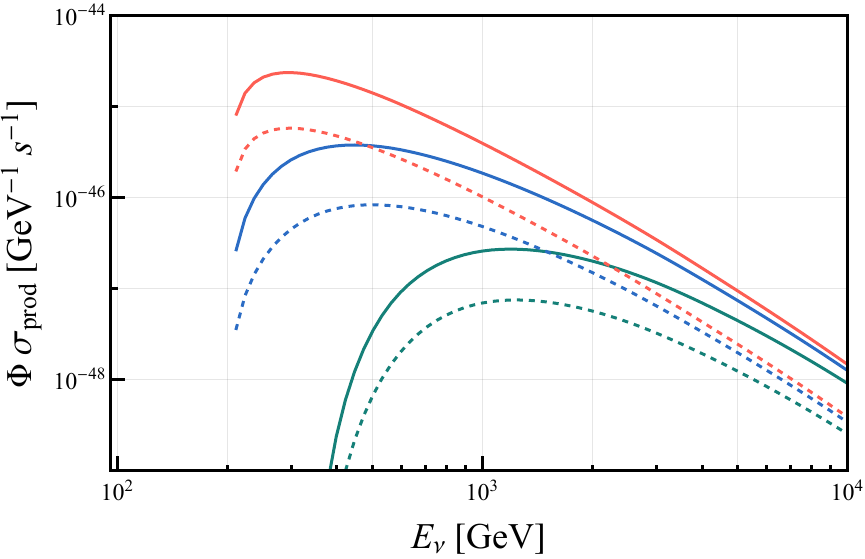}
    \end{subfigure}
    \caption{(Left) Cross section for the production singlet neutrinos $N$ as a function of $E_\nu$ $(|U_{N\ell}|^2 = 1)$. Solid and dashed curves correspond to neutrinos and antineutrinos in the initial state respectively. These cross sections are calculated for cuts of $E_1 > 100~\mathrm{GeV}$ on the hadronic cascade and $E_N > 100~\mathrm{GeV}$. (Right) Corresponding differential production rates, showing the range of atmospheric neutrino energies $E_\nu$ that dominate the production of $N$. In both the plots, the red, blue, and green curves correspond to $m_N = 1~\mathrm{GeV}, 10~\mathrm{GeV}$ and $20~\mathrm{GeV}$ respectively.}
    \label{fig: prod_cs}
\end{figure}

Singlet neutrinos are produced from the collisions of neutrinos with nucleons inside the detector, as illustrated in Fig.~\ref{fig: Nprod}. The differential cross section for this process is given by:
\begin{equation}
    \begin{split}
        \frac{d^2\sigma}{dx dQ^2} = \frac{\pi \alpha^2 |U_{N\ell}|^2}{8 s^4_W c^4_W}\sum_q \frac{f_{q}(x, Q^2)}{(Q^2 + m^2_Z)^2}&\left[(c_V + c_A)^2 \left(1 - \frac{m^2_N}{\hat{s}}\right) \right.\\
    &\left.+ (c_V - c_A)^2 \left(1 - \frac{Q^2}{\hat{s}}\right)\left(1 - \frac{Q^2}{\hat{s}} - \frac{m^2_N}{\hat{s}}\right)\right]    \; .   
    \end{split}
    \label{eq: diff_cs_parton}
\end{equation}
Here $\hat{s} = 2xm_n E_\nu$ is the square of the energy in the center of momentum frame, $c_V, c_A$ are the vector and axial couplings and $f_q(x, Q^2)$ denotes the parton distribution function (pdf) of the initial state quark $q = (u,d)$ inside the nucleon as a function of the DIS variables $(x, Q^2)$. Differential cross sections for the production through antineutrinos can be obtained by switching $c_V + c_A \leftrightarrow c_V - c_A$ in Eq.~(\ref{eq: diff_cs_parton}). We employ the proton pdfs from \texttt{nCTEQ15}~\cite{Kovarik:2015cma} for $f_q(x, Q^2)$. We estimate differential production rates using the all-direction averaged neutrino and antineutrino fluxes from~\cite{Honda:2015fha}. Production cross sections and differential production rates are shown as a function of $E_\nu$ for various values of $m_N$ in Fig.~\ref{fig: prod_cs}.

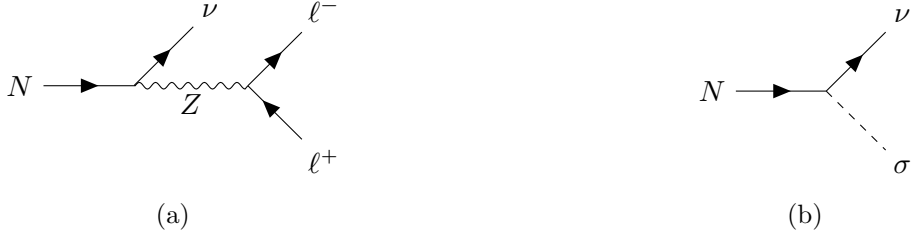
\begin{figure}[h!]
    \centering
\begin{subfigure}{0.45\textwidth}
    \centering
    \begin{tikzpicture}
    \begin{feynman}
    \vertex (i1) at (-2,0){$N$};
    \vertex (a) at (-0.5, 0);
    \vertex (f1) at (0.5, 1){$\nu$};
    \vertex (b) at (1,0);
    \vertex (f2) at (2, 1){$\ell^-$};
    \vertex (f3) at (2, -1){$\ell^+$};
    \diagram*{
    (i1)--[fermion](a),
    (a)--[fermion](f1),
    (a)--[boson,edge label' = $Z$, inner sep = 3pt](b),
    (b)--[fermion](f2),
    (b)--[anti fermion](f3),
    };
    \end{feynman};
    \end{tikzpicture}
    \caption{}
\end{subfigure}
\hfill
\begin{subfigure}{0.45\textwidth}
\centering
    \begin{tikzpicture}
    \begin{feynman}
    \vertex (a) at (-2,0){$N$};
    \vertex (i1) at (-0.5, 0);
    \vertex (b) at (0.5, 1){$\nu$};
    \vertex (c) at (0.5, -1){$\sigma$}; 
    \diagram*{
    (a)--[fermion](i1),
    (i1)--[fermion](b),
    (i1)--[scalar](c),
    };
    \end{feynman};
    \end{tikzpicture}
  \caption{}
\end{subfigure}
  \caption{Diagrams illustrating the decay modes of $N$. }
    \label{fig:N_decays}
\end{figure}

Once the singlet neutrinos have been produced, they can decay through two different modes (see Fig.~\ref{fig:N_decays}):
\begin{enumerate}
    \item [(i)] The interactions in Eq.~(\ref{eq: IR Lagrangian_1}) allow the $N$ to decay into SM particles through an off-shell $Z/W^\pm$, through processes such as $N \rightarrow \nu \ell^+ \ell^-, \nu \nu' \bar{\nu}', \nu q \bar{q}$, etc. Although these processes contain only a single factor of $U_{N\ell}$ at the amplitude level, the corresponding partial decay widths are additionally suppressed by a factor of $(m_N/v_{\rm EW})^4$,
    \begin{equation}
        \Gamma^{(3)}_N \approx \frac{m_N}{192 \pi^3} |U_{N\ell}|^2 \left(\frac{m_N}{v_\mathrm{EW}}\right)^4 \;.
    \label{eq: N_decay_SM}
    \end{equation}

    \item [(ii)] The $N$ can decay into $\sigma + \nu$ through the the $N-\nu-\sigma$ coupling in Eq.~(\ref{eq: IR Lagrangian_2}). This process is not suppressed by $v_{\rm EW}$, and the corresponding partial decay width is given by
    \begin{equation}
        \Gamma^{(2)}_N = \frac{y^2 m_N}{4 \pi^2} |U_{N\ell}|^2 \left( 1-\frac{m_\sigma^2}{m_N^2}\right)^2 .
    \label{eq: N_decay_HS}
    \end{equation}
\end{enumerate}

We focus on the region of parameter space where $m_N \ll v_{\rm EW}$ and $y \gtrsim \mathcal{O}(0.1)$ so that the two-body decays dominate. In this regime, the singlet neutrinos $N$ are short-lived and promptly decay into $\sigma + \nu$. The resulting $\sigma$ particles have two dominant decay modes at the tree level:
\begin{enumerate}
    \item [(i)] Four body decays mediated by an off-shell $W/Z$, with a rate that can be estimated as
    \begin{equation}
    \Gamma^{(4)}_\sigma \sim \frac{y^2 m_\sigma}{256 \pi^5} |U_{N\ell}|^4 \left(\frac{m_N}{v_{\rm EW}}\right)^4 \left(\frac{m_\sigma}{m_N}\right)^{10} \;.
    \label{four_body_decays}
    \end{equation}
    For $m_\sigma \gtrsim 1~\mathrm{GeV}$, we compute the width numerically using \texttt{MadGraph}. We include all possible final states, such as $\ell \ell' \nu \nu, \ell \nu q q' $, treating colored particles in the final state as partons. For $m_\sigma \lesssim 1~\mathrm{GeV}$, we perform a numerical integration including hadrons in the final state. We consider three- and four-body final states that include up to two hadrons such as $\overline{\nu_\ell}\ell^- \mathfrak{m}^+, \overline{\nu_\ell}\ell^- \mathfrak{m}^+\mathfrak{m}^0, \overline{\nu_\ell}\nu_\ell \mathfrak{m}^0, \overline{\nu_\ell}\nu_\ell \mathfrak{m}^+\mathfrak{m}^-$, where the $\mathfrak{m}$ are pseudoscalar mesons. Our treatment of hadronic decays follows an approach similar to that of Ref.~\cite{Bondarenko:2018ptm}, with the caveat that the final state contains an additional neutrino in our case. Details of our calculations may be found in Appendix~\ref{sec: NP_hadronic_decays}. The partial decay widths of the different hadronic decay modes are shown as a function of $m_N$ in Fig.~\ref{fig: sigma_hadronic_widths}. As shown here, for $m_\sigma \lesssim 1~\mathrm{GeV}$, $\sigma$ predominantly decays into $\overline{\nu_\ell}\nu_\ell \pi^0$.

    \begin{figure}[h!]
    \centering
    \begin{subfigure}{0.45\textwidth}
        \includegraphics[width=\linewidth]{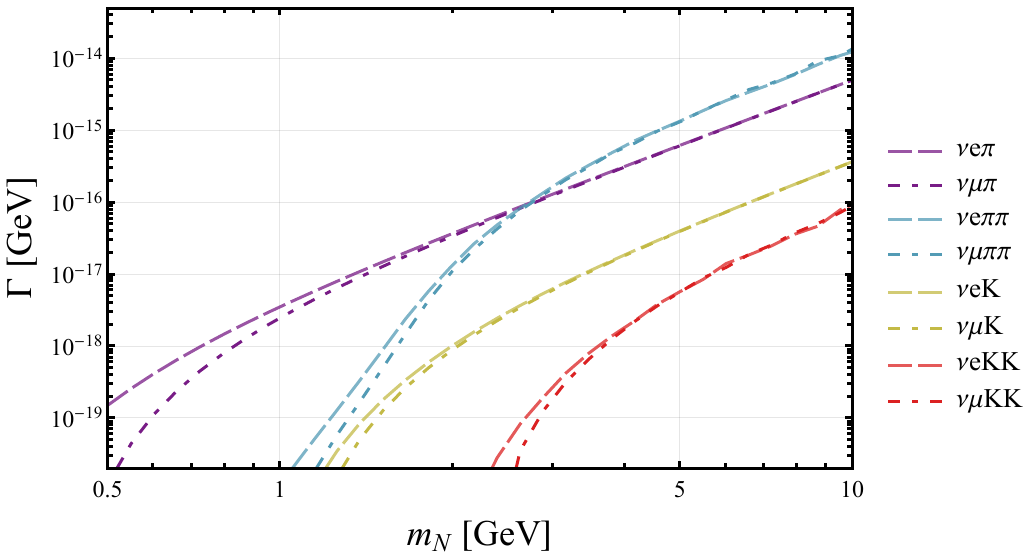}
    \end{subfigure}
    \hfill
    \begin{subfigure}{0.45\textwidth}
        \includegraphics[width=\linewidth]{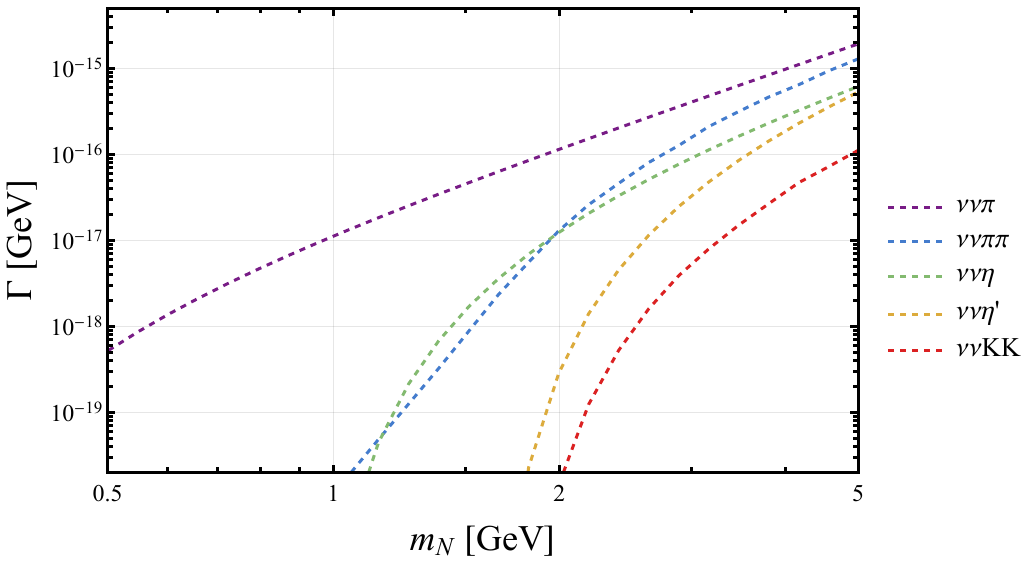}
    \end{subfigure}
    \bigskip
    \caption{Partial decay widths of the dominant hadronic decay modes of $\sigma$, mediated by the charged current (left) and the neutral current (right) interactions. Here, we set $y=1$, $|U_{N\ell}|^2=1$ and $m_\sigma/m_N = 0.6$.}
    \label{fig: sigma_hadronic_widths}
\end{figure}

    \item [(ii)] Two body decays $\sigma \rightarrow \nu \nu^c$, with rate given by
    \begin{equation}
    \Gamma^{(2)}_\sigma = \frac{y^2 m_\sigma}{8\pi} \left(\frac{m_\nu}{m_N}\right)^2 \;.
    \end{equation}
    Here we take the light neutrino masses to be of the order $m_\nu \sim 0.01~\mathrm{eV}$.
\end{enumerate}

\begin{figure}[h!]
  \centering
  \begin{subfigure}{0.45\textwidth}
    \centering
    \begin{tikzpicture}
    \begin{feynman}
      \vertex (i1) at (-1.75,  -1) {\(\sigma\)};
      \vertex (a)  at (-0.5, -1);
      \vertex (f1) at ( 0.5,  0) {\(\nu_{1}\)};
      \vertex (b)  at ( 1,  -1);
      \vertex (f2) at ( 2,  0) {\(\ell^{+}_{1}\)};
      \vertex (c)  at ( 2.5, -1);
      \vertex (f3) at ( 3.5, 0) {\(\bar{\nu}_{2}\)};
      \vertex (f4) at ( 3.5, -2) {\(\ell^{-}_{2}\)};
      \diagram*{
        (i1) -- [scalar] (a),
        (a)  -- [fermion] (f1),
        (a)  -- [anti fermion, edge label'={\(\bar{N}_{1}\)}, inner sep = 3pt]  (b)
              -- [anti fermion]  (f2),
        (b)  -- [boson, edge label'={\(W^{-}\)}] (c),
        (c)  -- [anti fermion] (f3),
        (c)  -- [fermion] (f4),
      };
    \end{feynman}
    \end{tikzpicture}
    \caption{}
    \label{fig: sigma_4}
  \end{subfigure}
  \hfill
  \begin{subfigure}{0.45\textwidth}
    \centering
    \begin{tikzpicture}
    \begin{feynman}
      \vertex (i1) at (-1.75, 0) {\(\sigma\)};
      \vertex (a)  at (-0.5, 0);
      \vertex (f1) at ( 0.5, 1) {\(\nu\)};
      \vertex (f2) at ( 0.5, -1) {\(\nu^c\)};
      \diagram*{
        (i1) -- [scalar] (a),
        (a)  -- [fermion] (f1),
        (a)  -- [anti fermion] (f2),
      };
    \end{feynman}
    \end{tikzpicture}
    \caption{}
    \label{fig: sigma_2}
  \end{subfigure}
  \par\vspace{2pc}
  \caption{Diagrams contributing to the two dominant decay modes of $\sigma$. Depending on the parameters $(m_N, |U_{N\ell}|^2)$, one of these two decay modes dominates. When $\sigma$ decays as in (b), it escapes detection. The region in the parameter space where invisible decays dominate is shown in Fig.~\ref{fig: sigma_lifetimes}.}
  \label{fig: sigma_decays}
\end{figure}
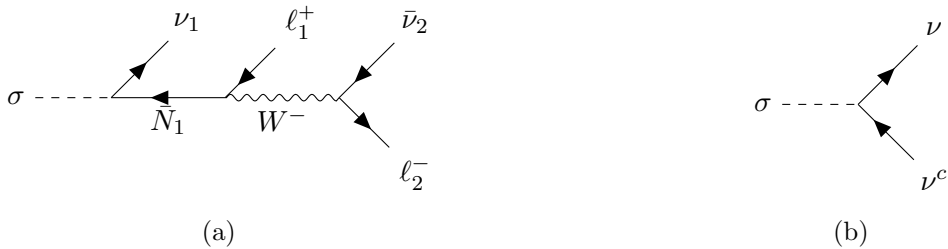

These tree level decay modes are illustrated in Fig.~\ref{fig: sigma_decays}. Depending on the values of $m_N$ and $|U_{N\ell}|^2$, one of these two decay modes dominates.

\begin{figure}[h!]
    \begin{subfigure}{0.3\textwidth}
    \centering
        \begin{tikzpicture}[scale= 0.4]
        \begin{feynman}
        \vertex (a) at (-4,0){$\sigma$};
        \vertex (i1) at (-2, 0);
        \vertex (i4) at (1,-2);
        \vertex(i5) at (1, 2);
        \vertex (b) at (3, -2){$\bar{\ell}$};
        \vertex (c) at (3, 2){${\ell}$};
        \diagram*{
        
        (a)--[scalar](i1),
        (i4)--[fermion,edge label = $\bar{N}/\bar{\nu}$](i1),
        (i1)--[fermion,edge label = ${N/\nu}$](i5),
        (b)--[fermion](i4),
        (i5)--[fermion](c),
        (i4)--[boson,edge label = $W$, inner sep = 3pt](i5),
        };
        \end{feynman};
        \end{tikzpicture}
    \caption{}
    \end{subfigure}
    \hfill
    \begin{subfigure}{0.3\textwidth}
    \centering
        \begin{tikzpicture}[scale=0.4]
            \begin{feynman}
                \vertex (a) at (-4,0){$\sigma$};
                \vertex (i1) at (-2,0);
                \vertex (i2) at (1,0);
                \vertex (i3) at (3,0);
                \vertex (i4) at (4.5, 2){${f}$};
                \vertex (i5) at (4.5, -2){$\bar{f}$};
                \diagram*{
                (a)--[scalar](i1),
                (i1)--[fermion, looseness = 0.5, half left,edge label = $N/\nu$](i2),
                (i2)--[fermion, looseness = 0.5, half left, edge label = $\bar{N}/\bar{\nu}$](i1),
                (i2)--[boson,edge label = $Z$](i3),
                (i3)--[fermion](i4),
                (i5)--[fermion](i3),
                };
            \end{feynman};
        \end{tikzpicture}
    \caption{}
    \end{subfigure}
    \hfill
    \begin{subfigure}{0.3\textwidth}
    \centering
        \begin{tikzpicture}[scale=0.4]
            \begin{feynman}
                \vertex (a) at (-4,0){$\sigma$};
                \vertex (i1) at (-2,0);
                \vertex (i2) at (1,0);
                \vertex (i3) at (3,0);
                \vertex (i4) at (4.5, 2){${f}$};
                \vertex (i5) at (4.5, -2){$\bar{f}$};
                \diagram*{
                (a)--[scalar](i1),
                (i1)--[fermion, looseness = 0.5, half left,edge label = $N$](i2),
                (i2)--[fermion, looseness = 0.5, half left, edge label = $\bar{\nu}$](i1),
                (i2)--[scalar,edge label = $h$](i3),
                (i3)--[fermion](i4),
                (i5)--[fermion](i3),
                };
            \end{feynman};
        \end{tikzpicture}
    \caption{}
    \end{subfigure}
    \caption{Loop processes that contribute to the two-body decay of $\sigma$ to SM fermions in the mass eigenbasis.}
    \label{fig:sigma_loop_decays}
\end{figure}
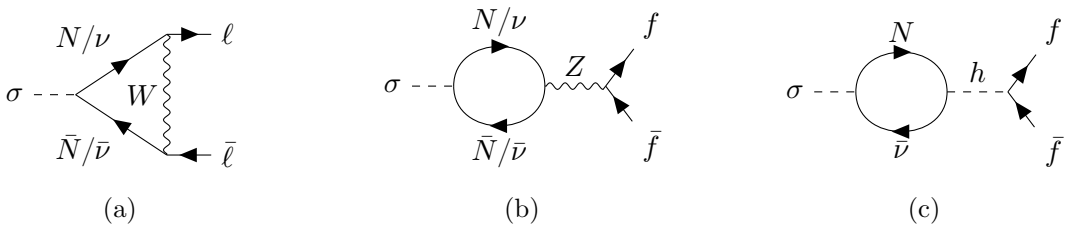

In addition to these tree level decay modes, $\sigma$ can decay to a pair of SM fermions through loop diagrams involving the $W$, $Z$ or Higgs $(h)$ as shown in Fig.~\ref{fig:sigma_loop_decays}. For each of the $W$ or $Z$ gauge bosons, there are three one-loop diagrams, corresponding to having $N$ and $\bar{N}$, $N$ and $\bar{\nu}$, or $\nu$ and $\bar{N}$ in the loop. (The contribution from the diagram with $\nu$ and $\bar{\nu}$ in the loop is suppressed by an additional factor of $(m_\nu/m_N)^2$ and can therefore be neglected). Although loop suppressed, the corresponding amplitudes arise at order $|U_{N\ell}|^2$, just as for the tree-level diagrams, and would naively be expected to contribute to the decay width at the same level. However, when the three diagrams are summed together, the total amplitude vanishes at order $1/v_{\rm{EW}}^2$ and the final result is of order $1/v_{\rm{EW}}^4$. In addition, the partial decay width to final state fermions of mass $m_f$ is suppressed by a factor of $m_f^2$ because of angular momentum considerations. The net contribution to the decay width is therefore suppressed compared to the naive expectation. The amplitude for the two-body decay to SM fermions through the mixing between $\sigma$ and $h$ that arises at loop level is also of the order $1/v_{\rm{EW}}^4$ and the contribution to the decay width again suppressed by a factor of $m_f^2$. 

From this discussion, it follows that the total contribution to the decay rate of $\sigma$ from these loop level processes can be estimated as 
 \begin{equation}
    \Gamma^{(\rm{loop})} \sim \frac{y^2|U_{N\ell}|^4 m_\sigma}{256 \pi^5} \left(\frac{m_N}{v_\mathrm{EW}} \right)^6 \left(\frac{m_f}{v_\mathrm{EW}}\right)^2\; .
\label{loop_decay_width}    
\end{equation}
 Comparing Eqs.~(\ref{loop_decay_width}) and (\ref{four_body_decays}), we see that the loop decays are highly suppressed compared to the four-body decays and can therefore be neglected. 

\begin{figure}[h!]
    \centering
    \begin{subfigure}{0.45\textwidth}
        \includegraphics[width=\linewidth]{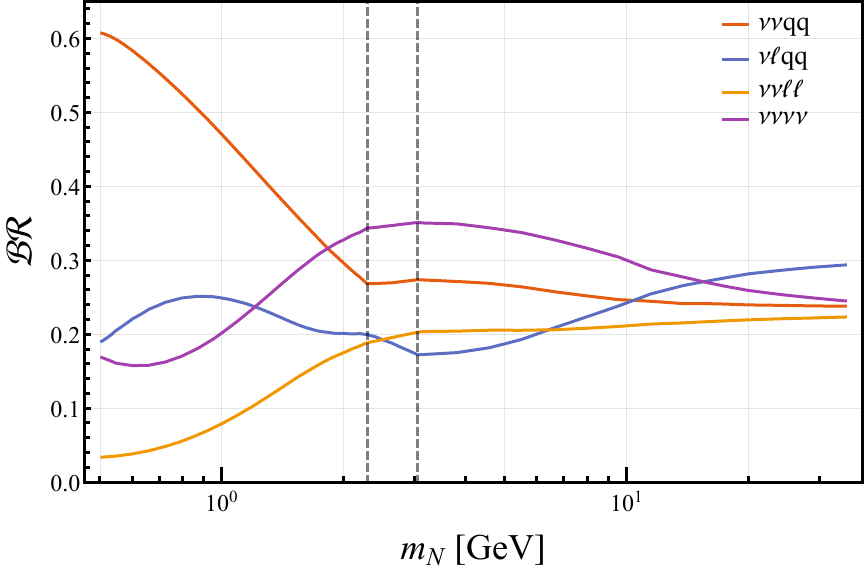}
    \end{subfigure}
    \hfill
    \begin{subfigure}{0.45\textwidth}
        \includegraphics[width=\linewidth]{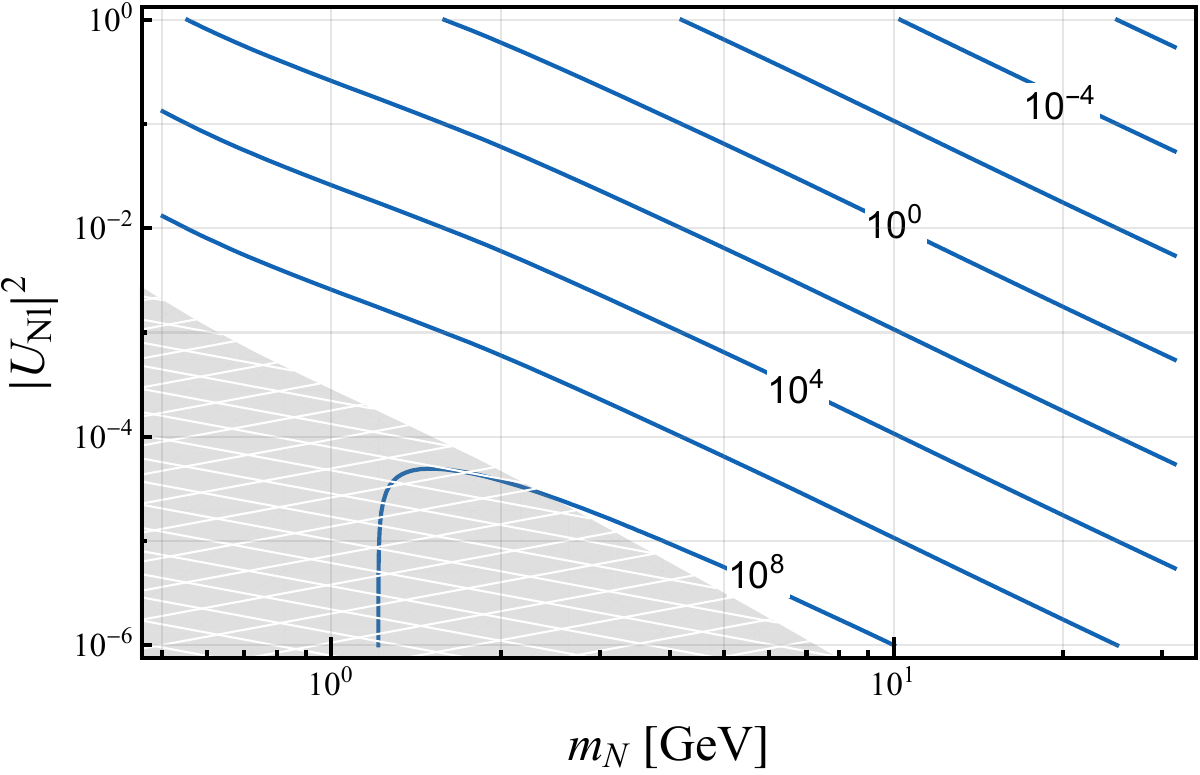}
    \end{subfigure}
    \bigskip
    \caption{(Left) Branching ratios of $\sigma$ to different final states, as a function of $m_N$. The dashed vertical lines show the transitions from hadronic calculations to the partonic ones for decays mediated by neutral-current and charged-current interactions, as discussed in Appendix~\ref{sec:hadronicdecays}. (Right) Contours of $c\tau_\sigma$ in meters as a function of $(m_N, |U_{N\ell}|^2)$. Here, we set $y=1$ and $m_\sigma/m_N = 0.6$. In the shaded region, the $\sigma$ predominantly decay invisibly $(\mathcal{BR}_\mathrm{inv} > 50\%)$. About $15-35\%$ of the four-body decays are invisible as can be read from the left plot.}
    \label{fig: sigma_lifetimes}
\end{figure}

Fig.~\ref{fig: sigma_lifetimes} shows the branching ratios and the decay length of $\sigma$ as a function of the parameters $(m_N, |U_{N\ell}|^2)$. It is clear that the $\sigma$ are long-lived and decay visibly in a large part of the parameter space. 
% As shown in Fig.~\ref{fig: sigma_lifetimes}, the $\sigma$ are long-lived and decay visibly in a large part of the parameter space. 

\subsection{Neutrino Induced Double Bang Event Rate at IceCube} 
\label{sec: NP_IceCube}

We are now in a position to determine the rate of double bang events at IceCube from this hidden sector model. The total event rate is given by:
\begin{equation}
    \mathcal{R}_{\mathrm{DB}} = 4\pi N_\mathrm{nuc} \int dE_\nu \Phi(E_\nu) \sigma_\mathrm{prod} (E_\nu) \bar{\epsilon} (E_\nu) \;,
\end{equation}
where $\Phi(E_\nu)$ is the differential flux of atmospheric neutrinos, $\sigma_{\rm prod}(E_\nu)$ represents the production cross section of singlet neutrinos and $N_\mathrm{nuc}$ is the number of nucleons inside the detector volume. The parameter $\bar{\epsilon}$ represents the average efficiency of observing a double bang signature. This depends on the geometry and sensitivity of the detector. It also depends on the lifetime and energy of the produced $\sigma$ particles, and the branching fraction of $\sigma$ into hadrons, electrons, and taus. We describe our calculation of $\bar{\epsilon}$ below.

It is necessary to apply a cut on the deposited energy $(E > E_\mathrm{min})$ in each cascade in order to register a signal in the digital optical modules (DOMs) and to suppress background rates. In addition, a cut needs to be placed on the spatial separation $(L > L_\mathrm{min})$ between the origin points of the two cascades in order to resolve the two cascades as separate. Ref.~\cite{Coloma_2017} proposed using $E_{\rm min} = 5$ GeV and $L_{\rm min} = 20$ m, while requiring the first cascade to be within a distance of $36~\mathrm{m}$ from at least 4 DOMs, corresponding to the in-ice SMT4 trigger. However, Ref.~\cite{Fischer:2024vmw} found that a sizable number of background events would survive after these cuts. In order to suppress the backgrounds, we apply very severe cuts of $E_\mathrm{min} = 100~\mathrm{GeV}$ on the deposited energies of both the cascades and $L_\mathrm{min} = 100~\mathrm{m}$. As shown in Ref.~\cite{Fischer:2024vmw}, reconstruction of the track length and the deposited energies is almost 100\% efficient for events passing these cuts. In addition, we performed a Monte Carlo simulation of the 86-string configuration of IceCube to verify that $E_\mathrm{min} = 100~\mathrm{GeV}$ results in a high (80\%) detection efficiency for the SMT4 trigger. Details of our Monte Carlo analysis may be found in Appendix~\ref{sec: DOM_Efficiency}. 

At energy depositions above $100~\mathrm{GeV}$, background events induced by the charged current interactions of muon neutrinos ($\nu_\mu-\mathrm{CC}$ interactions) and those induced by $\mu-\mathrm{CC}$ and $\mu-\mathrm{NC}$ interactions can be rejected due to the presence of muon tracks. Background events induced by $\nu_e-\mathrm{CC}$ and $\nu_\tau-\mathrm{CC}$ interactions do not pass the $L_\mathrm{min} = 100~\mathrm{m}$ cut. As a result, the dominant source of background events arises from two simultaneous hadronic cascades resulting from $\nu-\mathrm{NC}$ interactions. We estimate the annual rate of single hadronic cascades passing the $E_\mathrm{min}$ cut to be $N_0 = 6.5 \times 10^4~\mathrm{yr^{-1}}$. The annual rate of two such cascades occurring simultaneously within a time window of $\Delta t$ is given by $N_\mathrm{bkg} \simeq N^2_0 \left(\frac{\Delta t}{T}\right)$, where $T = 1~\mathrm{yr}$. Since the $\sigma$ are highly boosted in our signal, the two cascades would need to take place within a time window of $\Delta t_\mathrm{sig}\lesssim 1~\mathrm{km}/c \sim 3~\mathrm{\mu s}$. Considering a relatively longer time window of $\Delta t = 100~\mathrm{\mu s}$ to accommodate the finite buffer of the SMT triggers results in a background rate of $N_\mathrm{bkg} = 10^{-2}~\mathrm{yr^{-1}}$. We therefore expect less than one background event over the lifetime of the experiment.

Using the $(E_\mathrm{min}, L_\mathrm{min})$ cuts, the average efficiency $\bar{\epsilon}$ for events initiated by neutrinos of energy $E_\nu$ can be estimated as
\begin{equation}
    \bar{\epsilon} = \frac{1}{\sigma} \int dx dQ^2 \frac{d^2\sigma}{dx dQ^2} \prod\limits_{i=1, 2}\Theta\left(E^{i}_\mathrm{dep} - E_\mathrm{min}\right) \bar{\epsilon}_\mathrm{geom} \mathcal{BR}_\mathrm{vis} \;.
\end{equation}
Here $\mathcal{BR}_\mathrm{vis}$ is the branching ratio of $\sigma$ into hadrons, electrons, and taus, and $\bar{\epsilon}_\mathrm{geom}$ represents the geometric efficiency of the produced $\sigma$ to decay within the detector volume with a separation of at least $L_\mathrm{min}$ from the location of the first cascade. Since the neutrino flux is nearly uniform across the azimuthal angles, the singlet neutrinos $N$ are produced isotropically inside the detector volume. As shown in the right panel of Fig.~\ref{fig: prod_cs}, production of the hidden sector is dominated by the part of the neutrino spectrum where $E_\nu \gg m_N$. As a result, the $N$ are produced with a large boost, and with momenta essentially along the direction of the incoming neutrino. In order to estimate $\bar{\epsilon}_\mathrm{geom}$, we first run a Monte-Carlo simulation to obtain the probability distribution of the maximum track length $L_\mathrm{tr}$ inside the detector volume assuming the first bangs are uniformly distributed throughout the detector and that the direction of tracks is isotropically distributed. For a cubical detector of length $L_\mathrm{max}$, this distribution, denoted by $\mathcal{P}_\mathrm{tr}$ is linear in the range $L_\mathrm{tr} \in [0, L_\mathrm{max}]$ and dies away quickly for $L_\mathrm{tr} > L_\mathrm{max}$. Using this distribution, we then estimate $\bar{\epsilon}_\mathrm{geom}$ as
\begin{equation}
    \bar{\epsilon}_\mathrm{geom} = e^{-{L_\mathrm{min}}/{\gamma c\tau_\sigma}} -\int_{L_\mathrm{min}}^{L_\mathrm{max}} dL_\mathrm{tr} \mathcal{P}_\mathrm{tr} (L_\mathrm{tr}) e^{-{L_\mathrm{tr}}/{\gamma c\tau_\sigma}} \;.
\end{equation}

\begin{figure}[h!]
    \centering
    \begin{subfigure}{0.45\textwidth}
        \includegraphics[width=\linewidth]{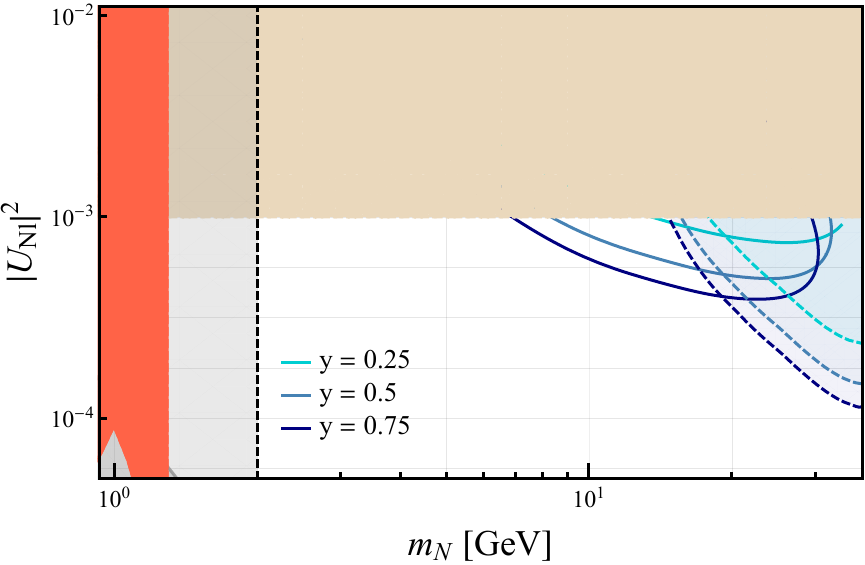}
    \end{subfigure}
    \hfill
    \begin{subfigure}{0.45\textwidth}
        \includegraphics[width=\linewidth]{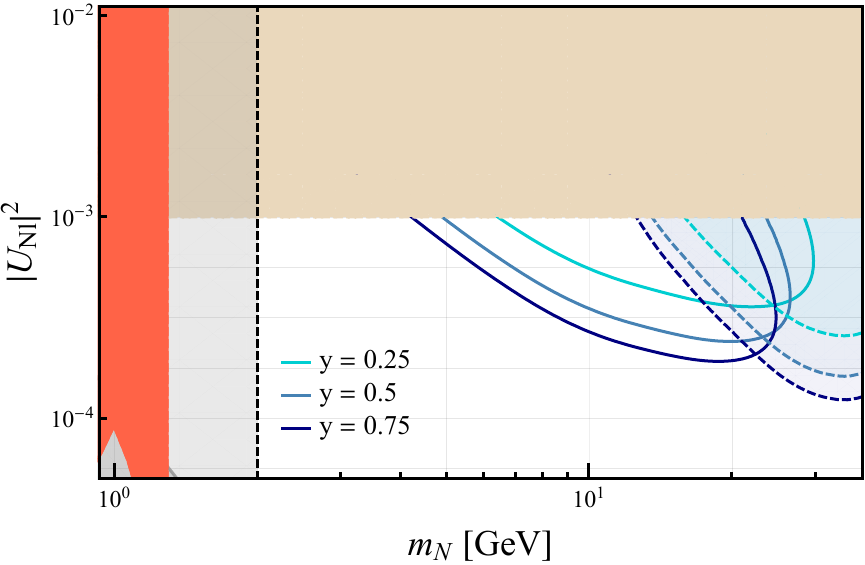}
    \end{subfigure}
    \caption{Contours of $N_\mathrm{evt} = 1$ for $m_{\sigma} / m_N = 0.6$ (left) and $m_{\sigma} / m_N = 0.8$ (right). The contours correspond to $y = 0.25,~0.5,~0.75$ respectively. The brown shaded region given by $|U_{N\ell}|^2 \gtrsim 10^{-3}$ is excluded by precision measurements. The shaded regions bounded by the dashed contours are excluded by searches for visible HNLs at the LHC. The region shaded in red corresponding to $m_N < m_\tau - 3m_\pi \sim 1.3~\mathrm{GeV}$ is excluded by the model independent search of BaBar. Additional constraints from searches for visible HNLs at beam dump experiments may be applicable in the gray shaded region corresponding to $m_N < m_D \sim 2~\mathrm{GeV}$. A detailed discussion of the current constraints is provided in Section~\ref{sec: NP_constraints}.}
    \label{fig: DB_N_evt}
\end{figure}

Contours corresponding to the number of events $N_\mathrm{evt} = 1$ over a period of fourteen years are shown in Fig.~\ref{fig: DB_N_evt} for the two benchmark values of $m_\sigma/m_N$. We see that in the mass range from about 4~GeV to about 20~GeV, IceCube has sensitivity to values of $|U_{N\ell}|^2$ well below the current bound of $10^{-3}$ from precision measurements. Furthermore, a part of the parameter space for $m_N \gtrsim 10~\mathrm{GeV}$ which is consistent with precision constraints has been ruled out by long-lived particle searches at the LHC. In Sec.~\ref{sec: NP_constraints} we discuss the details of these long-lived particle searches and other constraints on the model. Nevertheless, we see that IceCube has the potential to significantly improve on the current sensitivity to this class of hidden sector models in the mass range from a few GeV to about 20 GeV. 

\subsection{Current Constraints}
\label{sec: NP_constraints}

The singlet neutrinos $N$ inherit the couplings of the SM neutrinos to the $W$ and $Z$ gauge bosons, suppressed by a factor of the mixing $U_{N\ell}$. They can therefore be produced directly from the decays of on-shell $W$ and $Z$ gauge bosons, or from the leptonic and semi-leptonic decays of hadrons through an off-shell $W$ or $Z$. This being the case, the $N$ fall into the category of heavy neutral leptons (HNLs) from the viewpoint of searches at colliders and beam dumps. Existing bounds on HNLs from colliders and beam dumps that do not depend on how the $N$ decays can therefore be directly applied to this class of models. However, for HNL masses above a GeV, most searches search for specific visible decay modes arising from the HNL coupling to the weak gauge bosons of the SM. Since the corresponding bounds assume no additional invisible decay channels, these constraints need to be rescaled to determine the actual limits. 

For $m_N \lesssim 350~\mathrm{MeV}$, mixing angles of $|U_{N\ell}|^2 \gtrsim 10^{-8}$ are excluded based on the kinematics of the visible leptons in the leptonic decays of kaons~\cite{Yamazaki:1984sj}. A model-independent search by BaBar~\cite{BaBar:2022cqj} based on the momentum distribution of the pions arising from three-prong decays of tau leptons excludes $|U_{N\ell}|^2 \gtrsim 10^{-5}$ in the mass range given by $3m_\pi < m_N < m_\tau - 3 m_\pi$. Since these searches are independent of how the singlet neutrinos decay, the corresponding limits are directly applicable. However, for $m_N \gtrsim 1~\mathrm{GeV}$, most searches for HNLs at colliders and beam dumps are based on signals that require detection of the visible SM particles produced in their decays. Since, in our model, the singlet neutrinos primarily decay into a neutrino and a $\sigma$ rather than into visible SM final states, and the $\sigma$ is long-lived on detector time scales, these bounds are not directly applicable. However, constraints can still be placed on $|U_{N\ell}|^2$ based on the unitarity of the CKM matrix, as shown in~\cite{Bertoni:2014mva}. This excludes mixing angles $|U_{N\ell}|^2 \gtrsim 10^{-3}$. There are also lepton universality constraints based on ratios of different electroweak observables. However, these are proportional to factors of $(1 - |U_{Ni}|^2)/(1 - |U_{Nj}|^2)$, as shown in Table~II of~\cite{deGouvea:2015euy}. Since we are assuming flavor universality, $(U_{Ni} = U_{Nj})$, these bounds are not applicable. 

From Eqs.~(\ref{eq: N_decay_SM}) and~(\ref{eq: N_decay_HS}), the branching fraction for direct decays of $N$ into SM final states is given by
\begin{equation}
    \mathcal{BR}(N \rightarrow {\rm SM}) \approx \frac{1}{48\pi y^2}\left(\frac{m_N}{v_\mathrm{EW}}\right)^4 \left(1 - \frac{m^2_\sigma}{m^2_N}\right)^{-2}\; .
    \label{eq: NP_rescale}
\end{equation}
The current collider and beam dump constraints on $|U_{N\ell}|^2$ from searches for promptly decaying HNLs must therefore be rescaled by this factor to give the corresponding bounds in our case. Most relevant is the ATLAS search of Ref.~\cite{ATLAS:2019kpx}, which excludes mixing angles of $|U_{N\ell}|^2 \gtrsim 10^{-5}$ for promptly decaying HNLs in the mass range 7~GeV - 50~GeV. After rescaling by the factor shown in Eq.~(\ref{eq: NP_rescale}), this search excludes mixing angles of $|U_{N\ell}|^2 \gtrsim 3\times10^{-3}$ for $m_N = 50~\mathrm{GeV}$ and $y=0.1$. This is still weaker than the model-independent bound from CKM unitarity.

Given that the visible SM decay products of a $\sigma$ are identical to those of a singlet neutrino $N$, searches for long-lived HNLs at beam dumps and colliders are expected to constrain the parameter space of our model. Searches for visible HNLs produced in $D$-meson decays have been conducted by the CHARM~\cite{CHARM:1985nku, Boiarska:2021yho} and the BEBC WA66~\cite{WA66:1985mfx, Barouki:2022bkt} beam dump experiments. Both of these searches place constraints on HNLs that are lighter than $m_D \sim 2~\mathrm{GeV}$. In our case, IceCube is more sensitive than the precision measurements only in the mass range $m_N \gtrsim 4~\mathrm{GeV}$ (see Fig.~\ref{fig: DB_N_evt}), which is above the mass range constrained by CHARM and BEBC.

The CMS search~\cite{CMS:2023jqi} for visible long-lived HNLs produced in $W$ decays, based on a final state containing a prompt lepton $\ell_1$ from the decay $W \rightarrow \ell_1 + N$ and a displaced vertex arising from the subsequent decay $N \rightarrow \ell_2 q q$ is however relevant. The visible final states are the same as in our model, but the displaced vertex now arises from the decay of $\sigma$ rather than $N$. An important difference is that, in our case, the final state also contains two additional neutrinos: one from the decay of $N$ into $\sigma$ and another from the decay of $\sigma$. Moreover, this search categorizes signal events based on the significance of the transverse impact parameter $(d^{\mathrm{sig}}_{xy} = d_{xy}/\sigma_{xy})$ of the displaced lepton $\ell_2$. Since the decay lengths for our model are typically greater than hundreds of meters in the mass range of interest, the strongest constraints are set by the `very-displaced' regime, which is defined as $d^{\mathrm{sig}}_{xy} > 10$. The search excludes $N_\mathrm{evt} \geq 1$ in this regime. As described in~\cite{CMS:2014pgm}, $\sigma_{xy}$ is of the order $\sim 20~\mathrm{\mu m}$ and the track reconstruction efficiency drops to zero for displacements of the order $r_{DV} \sim 1 \mathrm{m}$. In order to estimate conservative bounds on our model, we generate events using \texttt{MadGraph} and apply the relevant cuts, assuming perfect energy resolution. This search requires the transverse momentum of the displaced jet to be $p^j_T > 20~\mathrm{GeV}$. Additionally, since the invariant mass of the prompt lepton and the displaced vertex is centered around $m_W$ in the case of regular HNLs, the search also requires $70~\mathrm{GeV} < m_{\ell_1 \ell_2 j} < 90~\mathrm{GeV}$. Due to the two additional neutrinos in our case, distributions of $p^j_T$ and $m_{\ell_1 \ell_2 j}$ are shifted towards lower values. As a reference, the $p^j_T$ and $m_{\ell_1 \ell_2 j}$ cuts roughly result in efficiencies of $15\%$ and $25\%$ respectively, for $m_N = 10~\mathrm{GeV}$ and $m_\sigma/m_N = 0.6$. For a given proper decay length $c\tau$, we calculate geometric efficiencies of the events by integrating over $d_{xy}>200~\mathrm{\mu m}$ and $r_{DV} < 1~m$. Using the average efficiency $\bar{\epsilon}$ (which includes the cuts and geometric factors), we exclude the region given by $N_\mathrm{evt} = \sigma_\mathrm{prod} \times \bar{\epsilon} \geq 1$.

The ATLAS search~\cite{ATLAS:2022atq} for visible HNLs based on leptonic decays resulting in a final state of one prompt lepton $\ell_1$ and two displaced leptons $\ell_2, \ell_3$ also places bounds on our model. This search excludes displaced vertices in the range $0.1~\mathrm{mm} < d_{xy} < 300~\mathrm{mm}$, using the reconstructed mass $m_{\rm HNL}$ of the HNL (assuming production from $W$ decay) to distinguish the signal from background. In our case, due to the two additional neutrinos in the final state, distributions of the reconstructed $m_{\rm HNL}$ are shifted towards lower values. As a result, the cut of $m_{\rm HNL} < 20~\mathrm{GeV}$ employed by this search leads to higher acceptance rates as compared to conventional HNLs of the same mass. Moreover, this search also requires the invariant mass of the particles from the displaced vertex $(m_{DV})$ to be greater than $2 - 5.5~\mathrm{GeV}$, with the exact value depending on the radial position of the displaced vertex $(r_{DV})$. We estimate conservative bounds on our model by applying the relevant cuts and integrating geometric factors in the range $0.1~\mathrm{mm} < d_{xy} < 300~\mathrm{mm}$ and $4~\mathrm{mm} < r_{DV}< 300~\mathrm{mm}$. Additionally, we assume the track reconstruction efficiency to be linearly decreasing with respect to $r_{DV}$ such that it is equal to $50\%$ at $r_{DV}=300~\mathrm{mm}$, as reported in~\cite{ATLAS:2017zsd}. 

\begin{figure}[h!]
    \centering
    \begin{subfigure}{0.45\textwidth}
        \includegraphics[width=\linewidth]{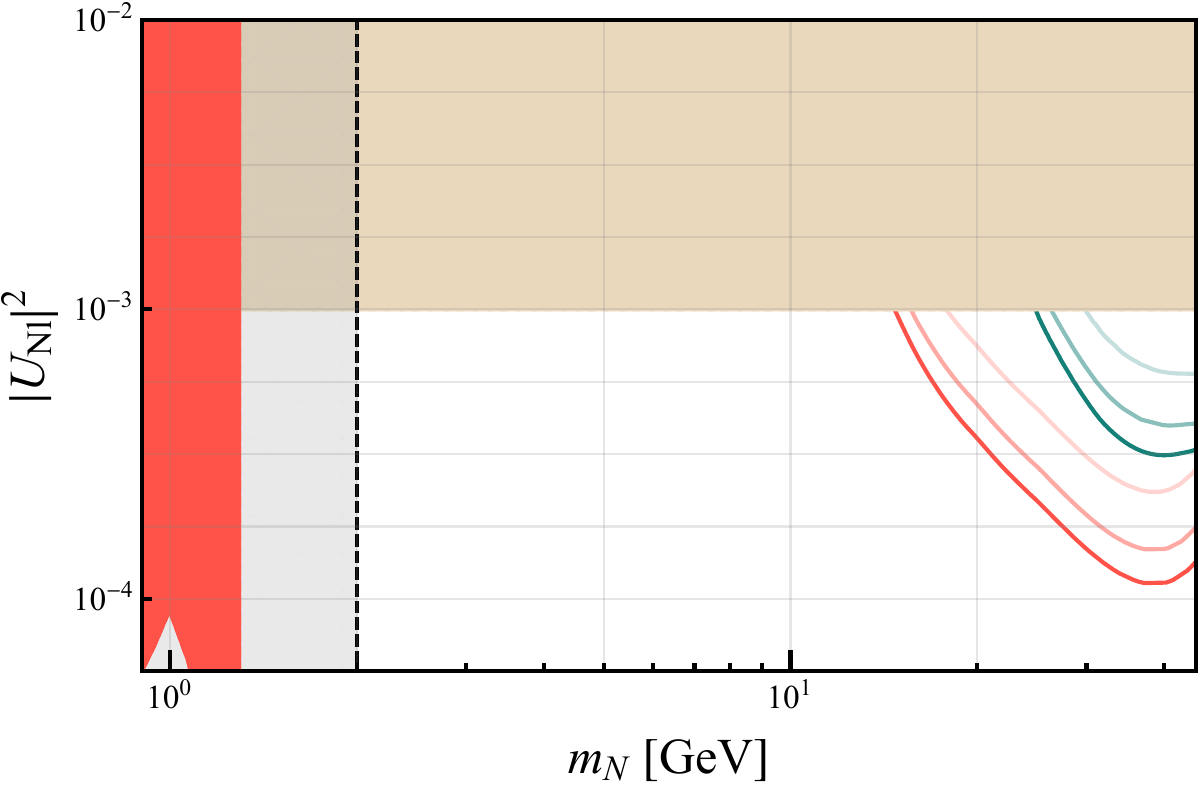}
    \end{subfigure}
    \hfill
    \begin{subfigure}{0.45\textwidth}
        \includegraphics[width=\linewidth]{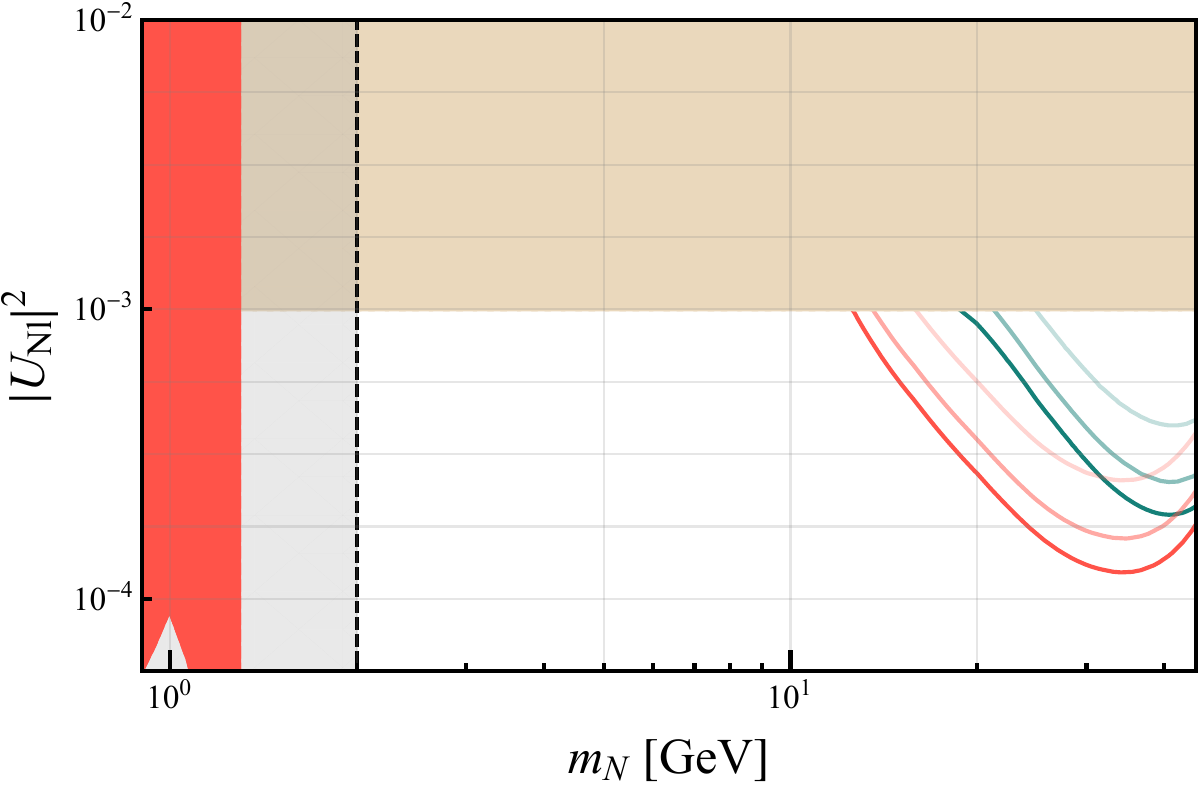}
    \end{subfigure}
    \caption{Regions of parameter space excluded by the CMS (green) and the ATLAS (red) searches for visible HNLs. The contours correspond to $y = 0.25, 0.5, 0.75$, in increasing levels of darkness. The shaded regions bounded by the dashed contours are excluded by searches for visible HNLs at the LHC. The region shaded in red corresponding to $m_N < m_\tau - 3m_\pi \sim 1.3~\mathrm{GeV}$ is excluded by the model independent search of BaBar. Additional constraints from searches for visible HNLs at beam dump experiments may be applicable in the gray shaded region corresponding to $m_N < m_D \sim 2~\mathrm{GeV}$.}
    \label{fig: LHC_constr_NP}
\end{figure}

Regions in the parameter space excluded by the CMS and ATLAS searches for the two benchmark values of $m_\sigma/m_N$ are shown in Fig.~\ref{fig: LHC_constr_NP}. Due to a higher track reconstruction efficiency, the ATLAS search results in stronger constraints on our model. The constraints become weaker for $m_N \lesssim 10~\mathrm{GeV}$ due to the $m_{DV}$ and $p^T_j$ cuts for the ATLAS and CMS searches respectively.

\section{The Hypercharge Portal}
\label{sec: HyperCharge Portal}

\subsection{Model Description}
\label{sec: HP_Model}
In this section, we consider a simplified model of a hidden sector that interacts with the SM through the hypercharge portal and gives rise to muon-induced double bang signals.  The hidden sector is assumed to contain a $U(1)_D$ dark gauge symmetry that is spontaneously broken. The vector boson associated with this broken gauge symmetry, which we label as $Z'$, mediates the portal interaction. The relevant part of the Lagrangian takes the form
\begin{equation}
\begin{split}
    \mathcal{L} \supset -\frac{1}{4} Z'^{\mu \nu} Z'_{\mu \nu}
    + \frac{1}{2} m^2_{Z'} Z'_\mu Z'^{\mu}
    + \frac{\epsilon}{2\cos{\theta_W}} Z'_{\mu \nu} B^{\mu \nu} \;.
\end{split}
\label{eq: HC_Lagrangian_1}
\end{equation}
In addition, the hidden sector contains two Dirac fermions, $\psi$ and $\chi$, which are both lighter than the $Z'$. We are interested in the case where $\psi$ is the heavier state and has charge 1 under $U(1)_D$, while $\chi$ is the LHSP and neutral under $U(1)_D$. The gauge coupling parameter in the hidden sector is denoted by $g_D$ and is assumed to be in the range $\mathcal{O}(0.1 - 1)$. We adopt the following Lagrangian for the fermions
\begin{equation}
\begin{split}
    \mathcal{L} \supset \bar{\psi} i D_\mu \psi + \bar{\chi} i \partial_\mu \chi
    - m_\psi \bar{\psi}\psi - m_\chi \bar{\chi}\chi - m\left(\bar{\psi}\chi + \bar{\chi}\psi\right).
\end{split}
\label{eq: HC_Lagrangian_2}
\end{equation}
Here $D_\mu \equiv \partial_\mu - i g_D Z'_\mu$ represents the covariant derivative. The mass parameter $m$ arises from a Yukawa coupling involving the hidden sector Higgs field that breaks the $U(1)_D$ gauge symmetry. This Yukawa coupling is assumed to be small enough so that $m\ll \Delta \equiv m_\psi - m_\chi$. For simplicity, the radial mode of the hidden sector Higgs field is taken to be significantly heavier than the gauge boson $Z'_\mu$ and the fermions $\psi$ and $\chi$, and it will not play any further role in our discussion.  

In the parameter region of interest, the fermions $\psi$ and $\chi$ are the mass eigenstates to a very good approximation. However, the mass mixing induces a small $\psi-\chi-Z'$ coupling,
\begin{equation}
\mathcal{L} \supset i g_D \frac{m}{\Delta} Z'_\mu \bar{\psi}\gamma^\mu \chi + \mathrm{h.c.}\; .
\label{eq: HC_Lagrangian_mixing}
\end{equation}
After diagonalizing the kinetic terms of $Z'_\mu$ and $B_\mu$, the $Z'$ acquires couplings to the SM fermions while the SM $Z$ acquires couplings to the hidden sector fermions. These interactions can be written as
\begin{equation}
    \mathcal{L} \supset -e \epsilon J^\mu_{\mathrm{em}} Z'_\mu + g_D \epsilon \tan{\theta_W} J'^\mu Z_\mu \;.
\label{eq: HC_portal_fermion_int}
\end{equation}

\begin{figure}[h!]
    \centering
  \begin{tikzpicture}[scale= 0.65]
   \begin{feynman}
   \vertex (a) at (-3,2){$\mu$};
  \vertex (b) at (3,2){$\mu$};
  \vertex (bb) at (3,1){$Z^\prime$};
   \vertex (i1) at (0,1);
   \vertex[blob] (i2) at (0,-1){};
   \vertex (i3) at (1.5, 1.5);
   \vertex (c) at (-3,-2){$n/p$};
   \vertex (d) at (3,-2){$n/p$};
  \diagram*{
  
  (i1)--[boson](i2),
   (a)--[fermion](i1),
   (i1)--[fermion](b),
   (i3)--[boson](bb),
   (c)--[fermion](i2),
   (i2)--[fermion](d),
  };
  \end{feynman};
  
  \end{tikzpicture}
    \caption{Production of $Z^\prime$ gauge bosons in muon-nucleon scattering.}
    \label{fig: Zprimeprod}
\end{figure}
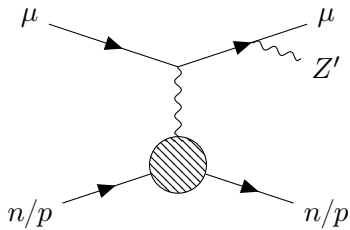

We now explain how this model can give rise to muon-initiated double bang signals at IceCube. Hidden sector states are predominantly produced through the emission of a $Z'$ in the muon-nucleon scattering process, as shown in Fig.~\ref{fig: Zprimeprod}. There is a large flux of downward-going muons produced in air showers initiated by cosmic rays in the atmosphere. In our analysis, we only consider $Z'$ gauge bosons that are produced outside the detector volume. There are two reasons for this. Firstly, although the low interaction rate of muons with matter allows them to travel all the way down to the IceCube detector, in the process they lose a significant portion of their energy. Consequently, the $Z'$ gauge bosons produced inside the IceCube detector are generally not very energetic, resulting in a low detection efficiency of the cascades from the muon-nucleon interaction. For reference, $\sim 90\%$ of the events result in $E_\mathrm{had} < 100~\mathrm{GeV}$ for $m_{Z'} = 5~\mathrm{GeV}$. Secondly, a visible muon track leading to the point where the cascade originates may lead to the event being vetoed. 

\begin{figure}[h!]
    \centering
    \includegraphics[width=0.7\linewidth]{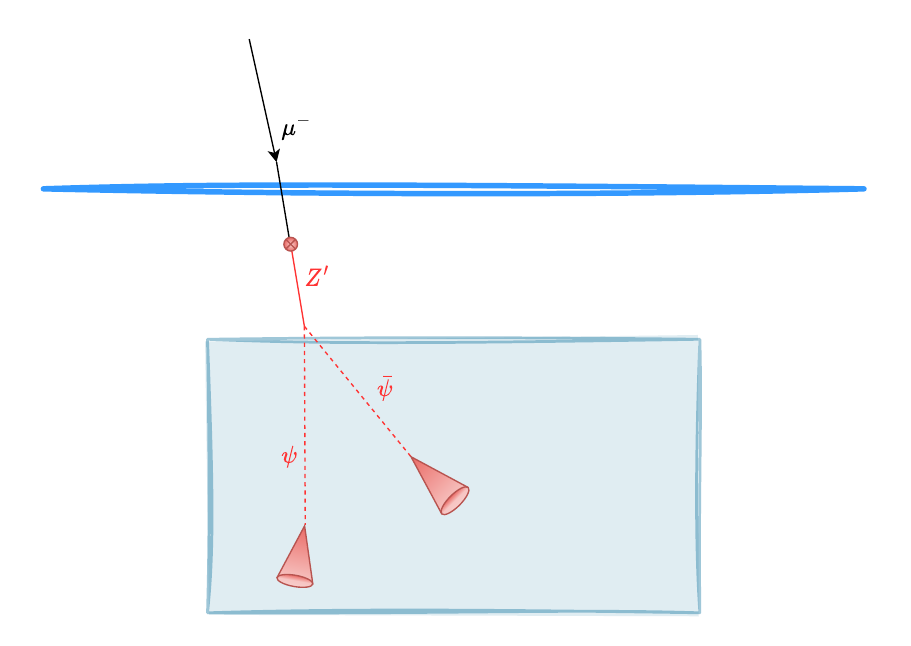}
    \caption{Illustration of the muon initiated double bang signature in the case of the hypercharge portal model. A $Z'$ is produced through muon-nucleon deep-inelastic scattering outside the instrumented volume. The $Z'$ promptly decays into $\psi\bar{\psi}$. Visible decays of $\psi$ and $\bar{\psi}$ inside IceCube result in the two cascades. The angle between the $\psi-\bar{\psi}$ tracks has been exaggerated for illustration purposes.}
    \label{fig: muon_DB_illustration}
\end{figure}

Since $g_D \sim \mathcal{O}(1)$, once a $Z'$ has been produced, it decays promptly to $\psi \bar{\psi}$. If $\psi$ and $\bar{\psi}$ are long-lived, with proper decay lengths of $c\tau \sim 10 - 1000~\mathrm{m}$ (see Section~\ref{sec: HP_prod_decay}), they can travel a significant distance before decaying into $\chi$ and SM particles. This then gives a rise to a double bang signal if both $\psi$ and $\bar{\psi}$ decay into final states containing hadrons, electrons, or taus within the detector volume. This process is illustrated in Fig.~\ref{fig: muon_DB_illustration}. The backgrounds for this signal are identical to those considered in Sec.~\ref{sec: NP_IceCube}, so we apply a very similar set of selection cuts to reduce them to zero.

As concrete realizations of this scenario, we study two benchmark sets of parameters. In these benchmarks, the values of $m$ and $g_D$ have been chosen to make the decay length of $\psi$ of the order of the physical dimensions of IceCube at the lower ($O(1)$~GeV) and higher ($O(10)$~GeV) ends of the range of interest for $m_{Z'}$. We fix the mass ratios of the fermions to the $Z'$. The benchmark points are defined in full in Table~\ref{tab: BPs_HC}.
%We determine the signal rate for these benchmark points as a function of $m_{Z'}$.

\begin{table}[h!]
    \centering
    \begin{tabular}{|c|c|c|c|c|}
    \hline
        & $m_\psi$ & $m_\chi$ & $m$ & $g_D$  \\
    \hline
    \hline
    BP1 & $m_{Z'}/4$ & $m_{Z'}/6$ & $m_{Z'}/200$ & $0.1$\\
    \hline
    BP2 & $m_{Z'}/4$ & $m_{Z'}/6$ & $m_{Z'}/40$ & $1.0$\\
    \hline
    \end{tabular}
    \caption{The two benchmark sets of parameters considered for the hypercharge portal model.}
    \label{tab: BPs_HC}
\end{table}

\subsection{Production and Decays of Hidden Sector Particles}
\label{sec: HP_prod_decay}
\begin{figure}[h!]
    \centering
    \begin{subfigure}{0.45\textwidth}
        \includegraphics[width=\linewidth]{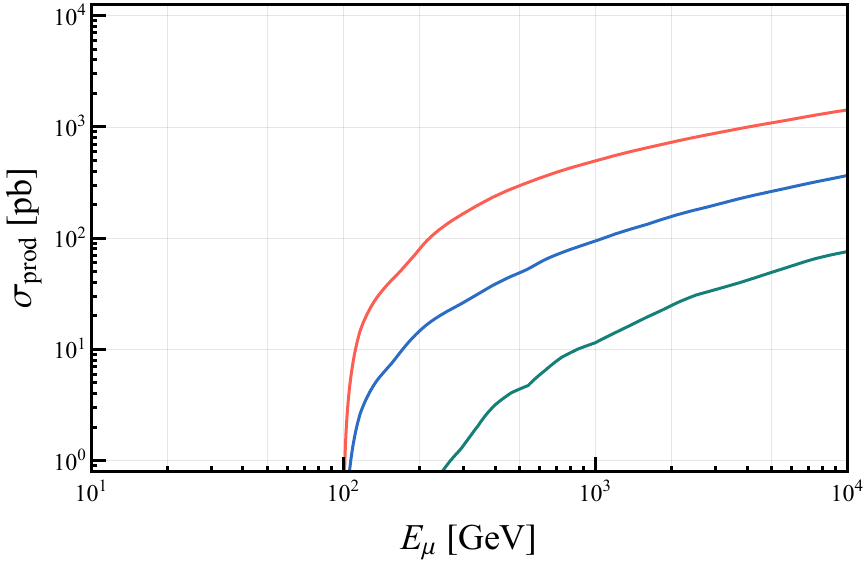}
    \end{subfigure}
    \hfill
    \begin{subfigure}{0.45\textwidth}
        \includegraphics[width=\linewidth]{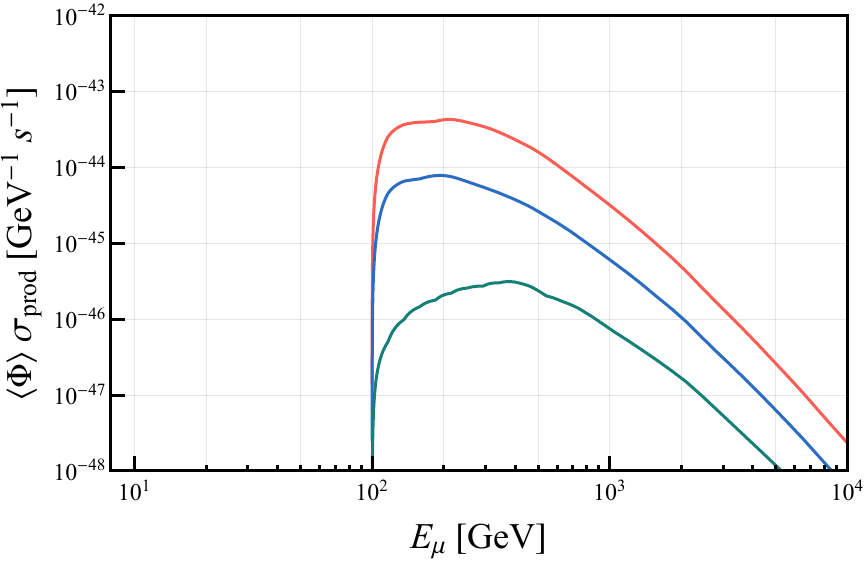}
    \end{subfigure}
    \caption{ (Left) Production cross section of the $Z'$ as a function of the incident muon energy $E_\mu$ (for $\epsilon = 1$). These cross sections are calculated for a cut of $E_1 > 100~\mathrm{GeV}$ on the hadronic cascade. (Right) The corresponding differential production rates. In both plots, the red, blue, and green curves correspond to $m_{Z'} = 1~\mathrm{GeV}, 5~\mathrm{GeV}$ and $10~\mathrm{GeV}$ respectively.}
    \label{fig: prod_cs_HC}
\end{figure}

The interaction in Eq.~(\ref{eq: HC_portal_fermion_int}) allows $Z'$ gauge bosons to be produced through radiation from initial or final state fermions in muon-nucleon or neutrino-nucleon scattering processes. Since the latter process is much less efficient at the relevant energies, we shall limit our attention to production from muon-nucleon scattering. We use the parametric form of the depth and angular dependence of the muon flux $\Phi(E_\mu, h, \theta)$ given in~\cite{Becherini:2005sr}. Production cross sections and differential production rates with the requirement that $E_{Z'} > 100~\mathrm{GeV}$ are shown as a function of the incident muon energy $E_\mu$ in Fig.~\ref{fig: prod_cs_HC}.

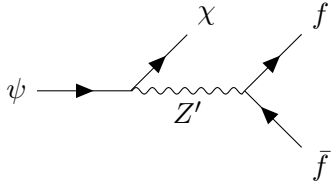
\begin{figure}[h!]
  \centering
    \begin{tikzpicture}
    \begin{feynman}
      \vertex (i1) at (-2,  0) {\(\psi\)};
      \vertex (a)  at (-0.5, 0);
      \vertex (f1) at ( 0.5,  1) {\(\chi\)};
      \vertex (b)  at ( 1,  0);
      \vertex (f2) at ( 2,  1) {\(f\)};
      \vertex (f3) at ( 2, -1) {\(\bar{f}\)};
      \diagram*{
        (i1) -- [fermion] (a),
        (a)  -- [fermion] (f1),
        (a)  -- [boson, edge label'={\(Z^\prime\)}]  (b),
        (b)  -- [fermion] (f2),
        (b)  -- [anti fermion] (f3),
      };
    \end{feynman}
    \end{tikzpicture}
\caption{Decay of $\psi$ through an off-shell $Z'$. Assuming that the hidden sector does not contain any other light degrees of freedom, the fermions $\psi$ decay as $\psi \rightarrow \chi f \bar{f}$, where the $f$ are SM fermions. These decays are visible in the cases when the $f$ are charged.}
\label{fig: psi_decay_visible}
\end{figure}

Since $g_D \gg e\epsilon$ in the parameter space of interest, the $Z'$ gauge bosons predominantly decay via $Z' \rightarrow \psi \bar{\psi}$. Once the fermions $\psi$ have been produced, they decay to a $\chi$ and SM fermions through an off-shell $Z'$, as shown in Fig.~\ref{fig: psi_decay_visible}. Due to the second term in Eq.~\ref{eq: HC_portal_fermion_int}, diagrams involving an off-shell $Z$ also contribute to these decays. However, compared to diagrams that involve an off-shell $Z'$, these are suppressed by a factor of $m^2_{Z'}/m^2_Z$ at the amplitude level, rendering their contributions negligible.

The decay width of the fermion $\psi$ can be estimated as
\begin{equation}
    \Gamma_\psi \sim \frac{e^2 \epsilon^2 g^2_D \Delta}{128 \pi^3}\left(\frac{m}{m_{Z'}}\right)^2 \left(\frac{\Delta}{m_{Z'}}\right)^2 \; .
\label{eq: HC_psi_decay_width}
\end{equation}
 For $\Delta \gtrsim 1~\mathrm{GeV}$, we determine the width numerically using \texttt{MadGraph}, with colored particles in the final state treated as partons. For $\Delta \lesssim 1~\mathrm{GeV}$, we perform a numerical integration including (up to two) hadrons in the final state. Further details may be found in Appendix~\ref{sec: HP_hadronic_decays}.

\begin{figure}[h!]
    \centering
    \begin{subfigure}{0.45\textwidth}
        \includegraphics[width=\linewidth]{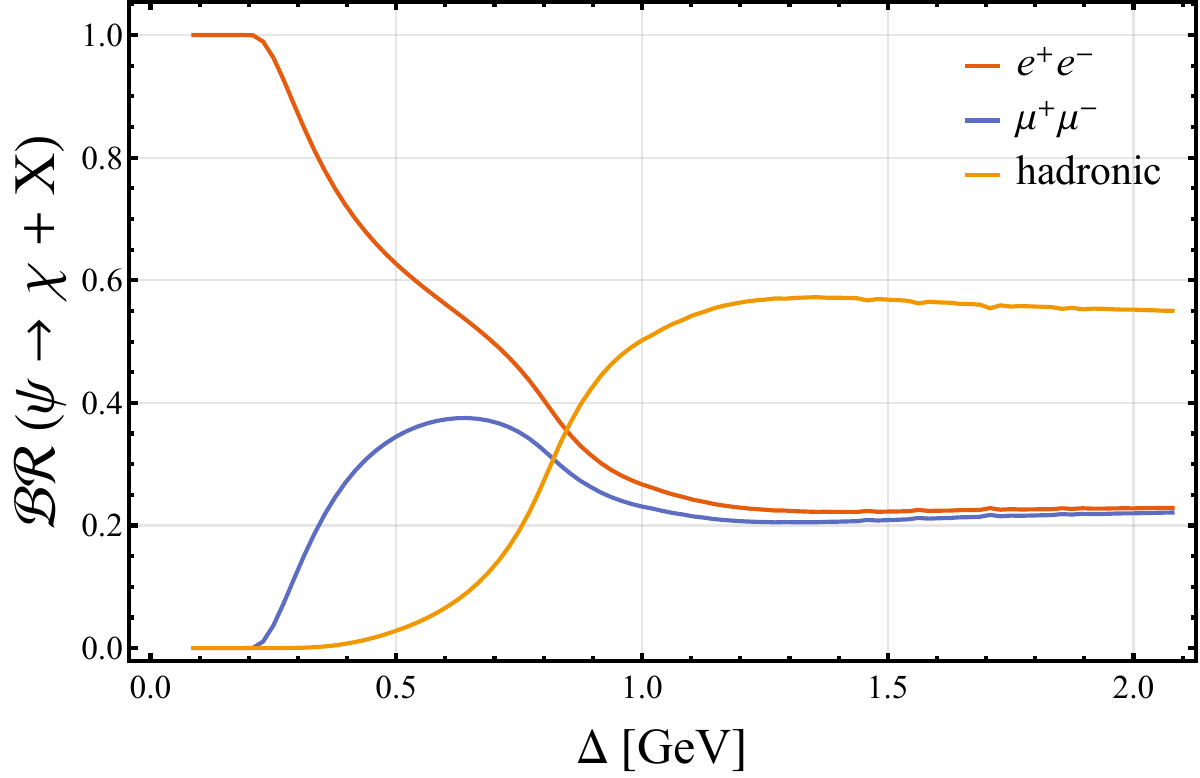}
    \end{subfigure}
    \hfill
    \begin{subfigure}{0.45\textwidth}
        \includegraphics[width=\linewidth]{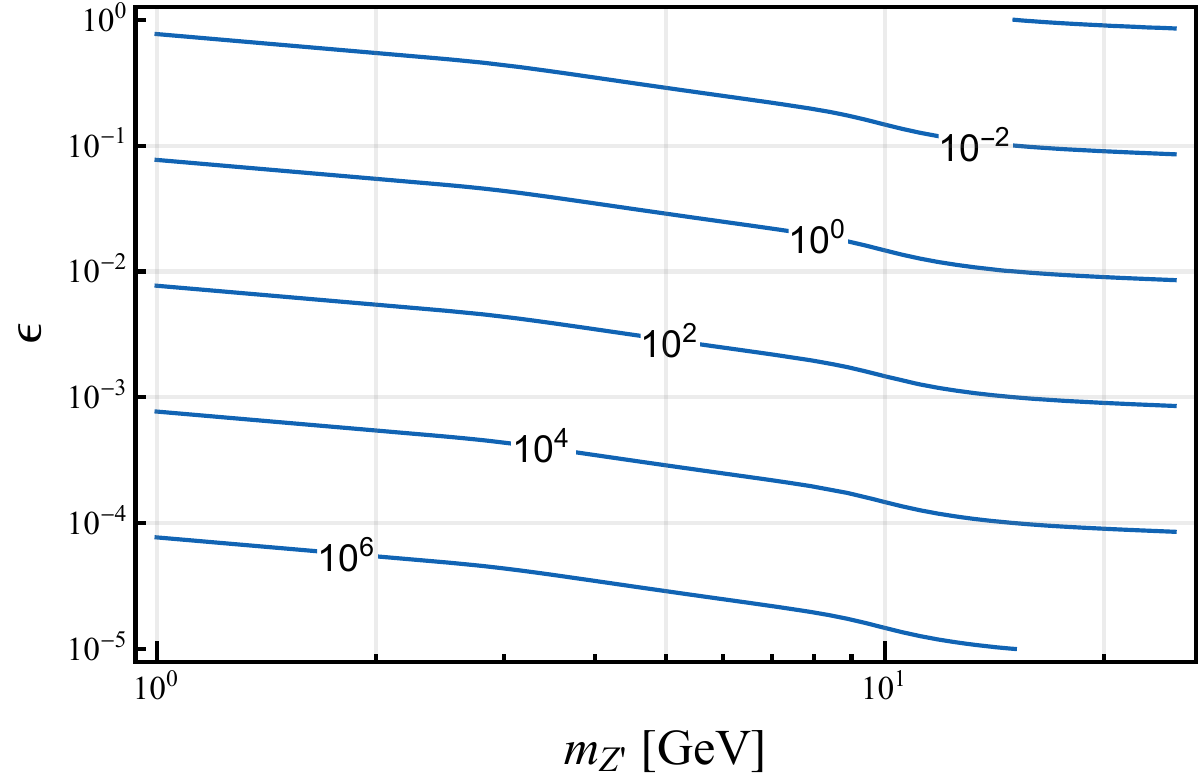}
    \end{subfigure}
    \caption{Left: Branching ratios of $\psi$ to different visible final states, as a function of the mass splitting $\Delta \equiv m_\psi - m_\chi$. Right: Contours of $c\tau_\psi$ in meters in the parameter space of $(m_{Z'}, \epsilon)$ for fixed $g_D = 0.1, ~m/m_{Z'} = 1/200$. In both the plots, $m_\psi/m_{Z'} = 1/4, ~m_\chi/m_{Z'} = 1/6$, consistent with our benchmarks.}
    \label{fig: HC_decay_plots}
\end{figure}

The left panel of Fig.~\ref{fig: HC_decay_plots} shows the branching ratios of $\psi$ into different SM final states as a function of the mass splitting $\Delta$. The right panel shows the proper lifetime $c\tau_\psi$ as a function of $m_{Z'}$ and $\epsilon$ for the first benchmark point. 

\subsection{Muon-Induced Double Bang Event Rate at IceCube}
\label{sec: HP_IceCube}
Next, we proceed to estimate the rate of muon-induced double bang events at IceCube. As explained in Section~\ref{sec: HP_Model}, we limit our attention to $Z'$ gauge bosons produced outside the detector volume. We take IceCube to be a cube of dimensions $x, y \in (-500, 500)~\mathrm{m}$ at a depth $z \in (-2450, -1450)~\mathrm{m}$. Then we sample the production of $Z'$ gauge bosons in a volume given by $x, y \in f_s(-500, 500)~\mathrm{m}$ and $z \in (-2450, 0)~\mathrm{m}$, where $f_s > 1$ is a scaling factor that determines the size of the simulated volume. Although a higher value of $f_s$ results in a larger target mass, tracks of $\psi$ and $\bar{\psi}$ are unlikely to pass through the detector volume if they are produced very far from the outer surface of the detector. We perform a Monte Carlo analysis by scanning over $f_s$ and estimating the rate of events for which both $\psi$ and $\bar{\psi}$ pass through the detector volume. We find that the rate of such events is saturated for $f_s = 1.25$, which we fix for the rest of the analysis. Additionally, we neglect events where the $Z'$ is produced inside the detector volume, since the visible muon track leading to the production point may lead to a veto.

With these assumptions, the rate of double bang events is given by
\begin{equation}
    \mathcal{R}_\mathrm{DB} = 4\pi N_\mathrm{nuc}\int dE_\mu \langle \Phi(E_\mu) \epsilon\rangle \sigma_\mathrm{prod}(E_\mu) \mathcal{BR}^2_\mathrm{vis}\; ,
\end{equation}
where $\langle.\rangle$ represents an average over production points $(x_0)$ and directions $(\theta, \phi)$ of the incoming muon and $\epsilon$ represents the detection efficiency. $\mathcal{BR}_\mathrm{vis}$ is the branching fraction of $\psi$ into final states containing electrons, taus and hadrons. The production points $x_0$ are sampled uniformly in the simulated volume. Since $E_\mu \gg m_{Z'}$ in the mass range of interest (see right panel of Fig.~\ref{fig: prod_cs_HC}), the $Z'$ gauge bosons are produced in the direction of the incoming muon. Consequently, both $\psi$ and $\bar{\psi}$ are also collinear with the incoming muon, although with different boost factors $\gamma_i$. So, for each event, we calculate the distances from $x_0$ at which the $\psi-\bar{\psi}$ track enters and exits the detector volume, denoted by $l_\mathrm{min}$ and $l_\mathrm{max}$ respectively. Furthermore, we require that the two cascades corresponding to the decays of $\psi$ and $\bar{\psi}$ are separated by a distance of at least $l_0 = 100~\mathrm{m}$. With these cuts, the geometric efficiency of a given event can be estimated as
\begin{equation}
    \epsilon_\mathrm{geom} = \int\limits_{x_1 = l_\mathrm{min}}^{l_\mathrm{max}} \frac{dx_1}{\gamma_1 c\tau} \int\limits_{x_2 = l_\mathrm{min}}^{l_\mathrm{max}} \frac{dx_2}{\gamma_2 c\tau} e^{-{x_1}/(\gamma_1 c\tau)} e^{-{x_2}/(\gamma_2 c\tau)} \Theta \left(|x_1 - x_2| - l_0\right)
\end{equation}
We also apply cuts on the energies of visible particles produced in the decays of $\psi$ and $\bar{\psi}$ to ensure that both cascades register DOM hits. We require that at least one of the two cascades deposit an energy greater than 100~GeV in the detector to pass the SMT4 trigger. We consider two different sets of cuts on the energy deposited by the second cascade, where we either require it to be greater than 50~GeV or greater than 100~GeV. The total efficiency of an event is given by the product $\epsilon = \epsilon_\mathrm{geom} \times \epsilon_\mathrm{cuts}$. We then average the product $\Phi(E_\mu; z, \theta)\times \epsilon$ over all the simulated events to obtain $\langle \Phi(E_\mu) \epsilon\rangle$. 

\begin{figure}[h!]
    \centering
    \begin{subfigure}{0.45\textwidth}
        \includegraphics[width=\linewidth]{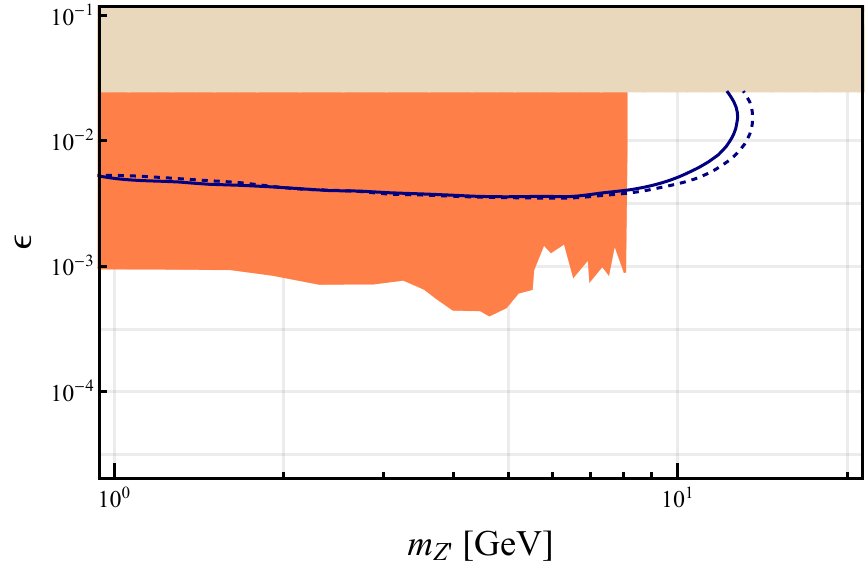}
    \end{subfigure}
    \hfill
    \begin{subfigure}{0.45\textwidth}
        \includegraphics[width=\linewidth]{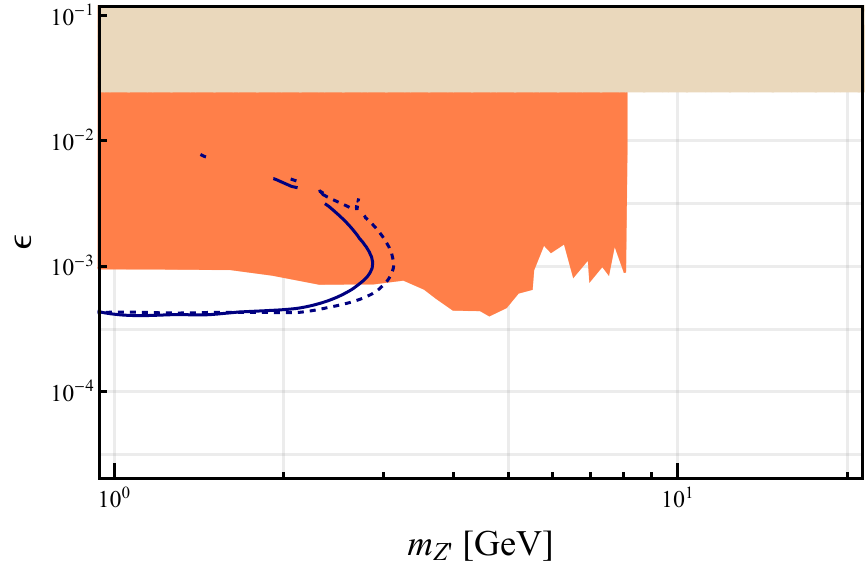}
    \end{subfigure}
    \bigskip
    \caption{Contours of $N_{\rm evt} = 1$ for BP1 (Left) and BP2 (Right). In both the plots, at least one of the two cascades is required to deposit an energy greater than 100~GeV. Dashed and solid contours correspond to cuts of 50~GeV and 100~GeV on the energy deposited by the second cascade. The shaded regions corresponding to $\epsilon \gtrsim 2.5\times10^{-2}$ and $m_{Z'} \lesssim 8~\mathrm{GeV}$ are excluded by precision electroweak measurements and BaBar respectively.}
    \label{fig: TB_N_evt_BP1}
\end{figure}

Contours corresponding to $N_\mathrm{evt} = 1$ over a period of fourteen years are shown in Fig.~\ref{fig: TB_N_evt_BP1} for both of our benchmark points. We see that in the parameter region of interest, IceCube has a reach extending beyond the current bounds from precision measurements and from the BaBar experiment. In Sec.~\ref{sec: HP_constraints} we discuss these and other constraints on the model in greater detail.

\subsection{Current Constraints}
\label{sec: HP_constraints}

The $Z'$ fits the general definition of a dark photon, and there are several experiments that have searched for invisible dark photons~\cite{Ilten:2018crw}. Beam dump experiments have placed very strong model-independent limits on dark photons with masses below a GeV. We shall therefore focus on the mass range $M_{Z'} \gtrsim 1 \; \rm{GeV}$, for which the strongest constraints are from collider searches. 
Invisible $Z'$ gauge bosons can be produced at $e^+e^-$ colliders, such as $\rm{LEP}$ and $\rm{BaBar}$, through the process $ e^+ e^- \to \gamma Z'$. These events give rise to a monophoton signature. The energy of the monochromatic photon can then be used to carry out a resonance search, leading to very strong limits \cite{Fox:2011fx, BaBar:2017tiz}. There are additional constraints from electroweak precision tests, since the mass of the SM $Z$ boson and its couplings to SM fermions are modified, but they are comparatively weak except when $M_{Z'}$ is close to $M_{Z}$ \cite{Curtin:2014cca}. In addition, since the $Z$ boson will also couple to the hidden sector fermions, see Eq.~(\ref{eq: HC_portal_fermion_int}), a precision measurement of the invisible branching fraction of the $Z$ boson also places a bound on the mixing parameter $\epsilon$, but this constraint is relatively mild. This bound as well as the one obtained by the BaBar experiment are shown in figure~\ref{fig: TB_N_evt_BP1}.

Although the $Z'$ decays invisibly in our case, visible signatures are possible at beam dumps and colliders when one or both of the $\psi$ decay within the detector volume. Due to the large proper decay lengths of $\psi$ in the parameter space of interest, searches for displaced vertices can potentially constrain our model. As described in~\cite{Ilten:2018crw}, the $Z'$ is produced in the decays of $\pi^0, \eta, \eta'$ mesons at proton beam dump experiments such as CHARM and PS191. As a result, constraints from visible searches at beam dump experiments are only applicable in the mass range $m_{Z'} < m_{\eta'} \sim 0.95~\mathrm{GeV}$, which is below our mass range of interest.

At the LHC, the $Z'$ is most abundantly produced in association with a jet $(p p \rightarrow Z' j)$ at the primary vertex. With the requirement that $p^j_T > 20~\mathrm{GeV}$ at the primary vertex, the production cross section for this process scales as $\sigma_\mathrm{prod} \sim \epsilon^2 \times 10^{5}~\mathrm{pb}$ in the mass range of . Displaced vertices arising from the decays of $\psi$ are either composed of pairs of leptons or hadronic jets. The ATLAS search of~\cite{ATLAS:2022gbw} places constraints on displaced vertices formed by hadronic jets for proper decay lengths as large as $c\tau \sim 72~\mathrm{m}$. However, this search applies a cut of $p_T > 30~\mathrm{GeV}$ on the displaced jets, resulting in negligible signal efficiencies in our case. The ATLAS search of~\cite{ATLAS:2018rjc} places constraints on displaced di-muon vertices in the muon spectrometer. The cut on the invariant mass of the displaced vertex defined as $m_{\mu\mu} > 15~\mathrm{GeV}$ completely eliminates the signal in our case, since for our benchmark points the mass splitting $\Delta = m_{Z'}/12 \ll 15~\mathrm{GeV}$. As a result, this search also does not constrain the model.

\section{Conclusion}
\label{sec: Conclusion}

In this paper, we have explored the possibility of directly detecting hidden sector particles at the IceCube neutrino telescope. Hidden sector particles can be produced through the collision of an energetic neutrino or atmospheric muon with a nucleon. Once produced, they decay down to the lightest particle in the hidden sector, the LHSP. If the LHSP is long-lived, this scenario can give rise to distinctive double bang events, characterized by two separate cascades. 

We have investigated IceCube's sensitivity to hidden sectors that interact with the SM through the neutrino and hypercharge portals. For each portal, we have assessed IceCube's reach using a simplified model of a hidden sector that gives rise to a double bang signal. For the case of the neutrino portal, we have considered events initiated by an incoming neutrino, whereas for the hypercharge portal, we have considered events initiated by an incoming muon. 

In our analysis, we have primarily focused on events in which the two cascades are spatially separated by at least 100 m and each deposits at least 100 GeV of energy. We assume that a suitable trigger can be developed for such events. Apart from possible instrumental backgrounds, which we do not attempt to estimate, these cuts should result in a negligible rate of background events. With these assumptions, we find that IceCube has the potential to significantly improve on the current sensitivity to hidden sector particles in the mass range from about a GeV to about 20 GeV.

\acknowledgments
SA and ZC are supported by the National Science Foundation under Grant Number PHY-2210361. The research of CK and RPRS is supported by the National Science Foundation Grant Number PHY-2210562. ZC would like to thank Subir Sarkar and Greg Sullivan for useful discussions.

\appendix
\makeatletter
\renewcommand{\@seccntformat}[1]{Appendix \csname the#1\endcsname\quad}
\makeatother
\renewcommand{\thesection}{\Alph{section}}
\numberwithin{equation}{section}

\section{Hadronic Decays of $\sigma$ and $\psi$}
\label{sec:hadronicdecays}

For the two portals that we consider, the LHSP has decay modes containing quark-antiquark pairs in the final state. When the mass of LHSP is well above 1~GeV, the hadronic decay widths are well approximated by the parton-level calculations. However, a reasonable part of our mass range lies close to or below 1~GeV, where we need to consider decays into hadrons. The authors of~\cite{Bondarenko:2018ptm} have calculated the hadronic decay widths in the case of conventional HNLs. We adopt a similar strategy in our calculations, employing the hadronic matrix elements, decay constants, and form factors given in the appendix of Ref.~\cite{Bondarenko:2018ptm}.

\subsection{Neutrino Portal}
\label{sec: NP_hadronic_decays}

There are two decay modes of $\sigma$ that contain quark-antiquark pairs in the final states: decays of the form $\sigma \rightarrow \nu \ell u\overline{d}$ mediated by charged-current interactions and decays of the form $\sigma \rightarrow \nu \nu q \overline{q}$, mediated by neutral-current interactions. In both cases, we limit ourselves to final states that contain one or two pseudoscalar mesons, resulting in three-body and four-body decay modes, respectively. We list the squared matrix elements for each of the decay modes below, where we denote the common factor by $\kappa^2 \equiv y^2|U_{N\ell}|^4$ for readability. Additionally, the results shown below contain factors for multiple charge assignments in the final state ($\nu\ell\mathfrak{m} \equiv \overline{\nu_\ell}\ell^- \mathfrak{m}^+, \nu_\ell\ell^+ \mathfrak{m}^-$) and for sums over lepton flavors whenever the result is independent of the charged lepton mass. The labeling of the $m^2_{ij}$ is based on the order in which the final-state particles appear on the left hand side of each equation.

\begin{eqnarray}
    \overline{|\mathcal{M}|^2}(\sigma \rightarrow \nu\ell \mathfrak{m}) &=& \frac{4\kappa^2|V_{ij}|^2 f^2_\mathfrak{m}G^2_F m^2_N}{(m^2_{23} - m^2_N)^2}\times\nonumber\\
    &&\left[(m^2_{13} - m^2_\mathfrak{m})(m^2_{23} - m^2_\mathfrak{m} -m^2_\ell) - m^2_\mathfrak{m}(m^2_{12} - m^2_\ell)\right]\; .\\
    \overline{|\mathcal{M}|^2}(\sigma \rightarrow \nu\ell \mathfrak{m}\mathfrak{m}) &= &\frac{8\kappa^2|V_{ij}|^2 |F_\mathfrak{m}(m^2_{34})|^2 G^2_F m^2_N}{(m^2_{234} - m^2_N)^2} \times\nonumber\\
        &&\left[(m^2_{13} - m^2_{14})(m^2_{23} - m^2_{24}) - (4m^2_\mathfrak{m} - m^2_{34})(m^2_{12} - m^2_\ell)\right]\;.\\
\overline{|\mathcal{M}|^2}(\sigma \rightarrow \nu\nu \mathfrak{m}) &= &\frac{3\kappa^2f^2_\mathfrak{m}G^2_F m^2_N (m^2_{13}+m^2_{23}-2m^2_N)^2}{(m^2_{13} - m^2_N)^2(m^2_{23} - m^2_N)^2} \times\nonumber\\
&&\left[(m^2_{13} - m^2_\mathfrak{m})(m^2_{23} - m^2_\mathfrak{m})- m^2_\mathfrak{m}m^2_{12}\right]\;.\\
\overline{|\mathcal{M}|^2}(\sigma \rightarrow \nu\nu \mathfrak{m}\mathfrak{m}) &= & \frac{6\kappa^2(1-s^2_W)^2|F_\mathfrak{m}(m^2_{34})|^2G^2_F m^2_N}{(m^2_{134} - m^2_N)^2(m^2_{234} - m^2_N)^2} (m^2_{134}+m^2_{234}-2m^2_N)^2 \times\nonumber\\
       & &\left[(m^2_{13} - m^2_{14})(m^2_{23} - m^2_{24}) - (4m^2_\mathfrak{m} - m^2_{34})m^2_{12}\right]\;.
\end{eqnarray}

\begin{figure}[h!]
    \centering
    \begin{subfigure}{0.45\textwidth}
        \includegraphics[width=\linewidth]{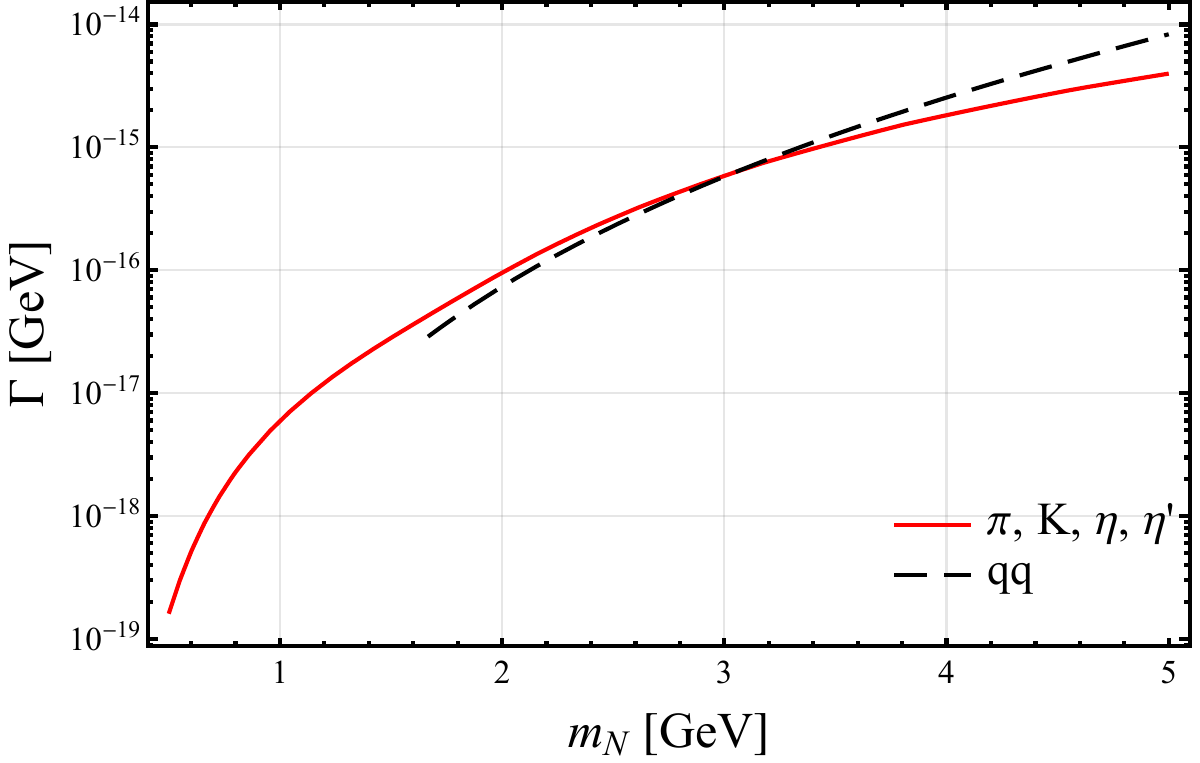}
    \end{subfigure}
    \hfill
    \begin{subfigure}{0.45\textwidth}
        \includegraphics[width=\linewidth]{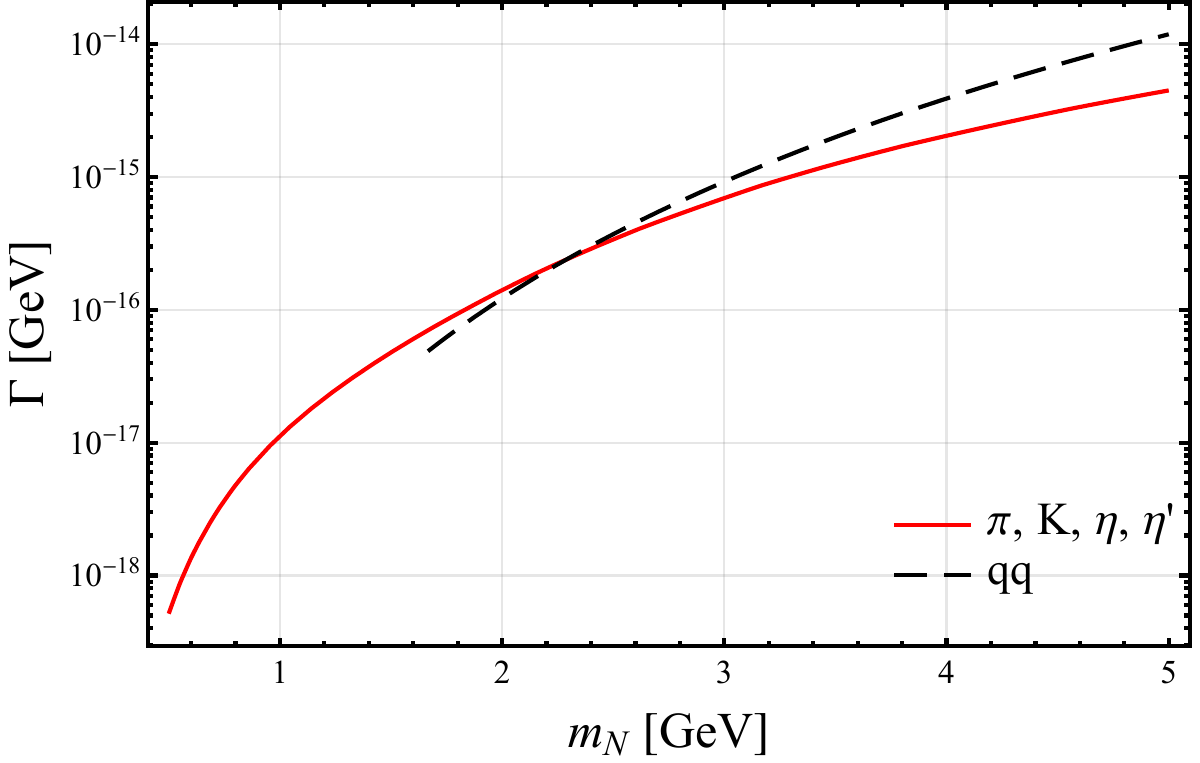}
    \end{subfigure}
    \bigskip
    \caption{Comparison of the decay widths obtained using the hadronic calculation and the parton-level calculation for $m_\sigma/m_N = 0.6$. The left and right panels show the decays mediated by charged-current and neutral-current interactions respectively.}
    \label{fig: NP_compare_hadronic}
\end{figure}

In the case of four-body decays, we decompose the phase space into three sequential two-body phase spaces, resulting in the following form for the differential decay width,
\begin{equation}
    d\Gamma = \frac{\overline{|\mathcal{M}|^2}\lambda^{1/2}_1\lambda^{1/2}_2\lambda^{1/2}_3}{2^{18}\pi^8 M^3 m^2_{234}m^2_{34}}dm^2_{234}dm^2_{34} d\Omega_1d\Omega_2d\Omega_3 \; ,
\end{equation}
where $M$ is the parent mass, $\lambda_1 = \lambda(M^2, m^2_1, m^2_{234})$, $\lambda_2 = \lambda(m^2_{234}, m^2_2, m^2_{34})$, $\lambda_3 = \lambda(m^2_{34}, m^2_3, m^2_{4})$, and $\Omega_i$ for $i=1, 2, 3$ are the solid angles in the rest frames of $M, m_{234}, m_{34}$, respectively. $\lambda(a, b, c) = a^2 + b^2 + c^2 - 2(ab + bc+ ca)$ is the K\"all\'en function. Then we perform a Monte Carlo integration by sequentially sampling $m^2_{234} \in [(m_2 + m_3+m_4)^2, (M-m_1)^2]$ and $m^2_{34} \in [(m_3 + m_4)^2, (m_{234}-m_2)^2]$, while $\Omega_i$ are independently and uniformly sampled. A comparison between the hadronic calculation and the parton-level calculation is shown in Fig.~\ref{fig: NP_compare_hadronic}. We treat decays mediated by charged-current and neutral-current interactions independently, and in both the cases, transition from the hadronic calculation to the parton-level calculation at the value of $m_N$ where the two calculations match.

\subsection{Hypercharge Portal}
\label{sec: HP_hadronic_decays}

\begin{figure}[h!]
    \centering
    \includegraphics[width=0.5\linewidth]{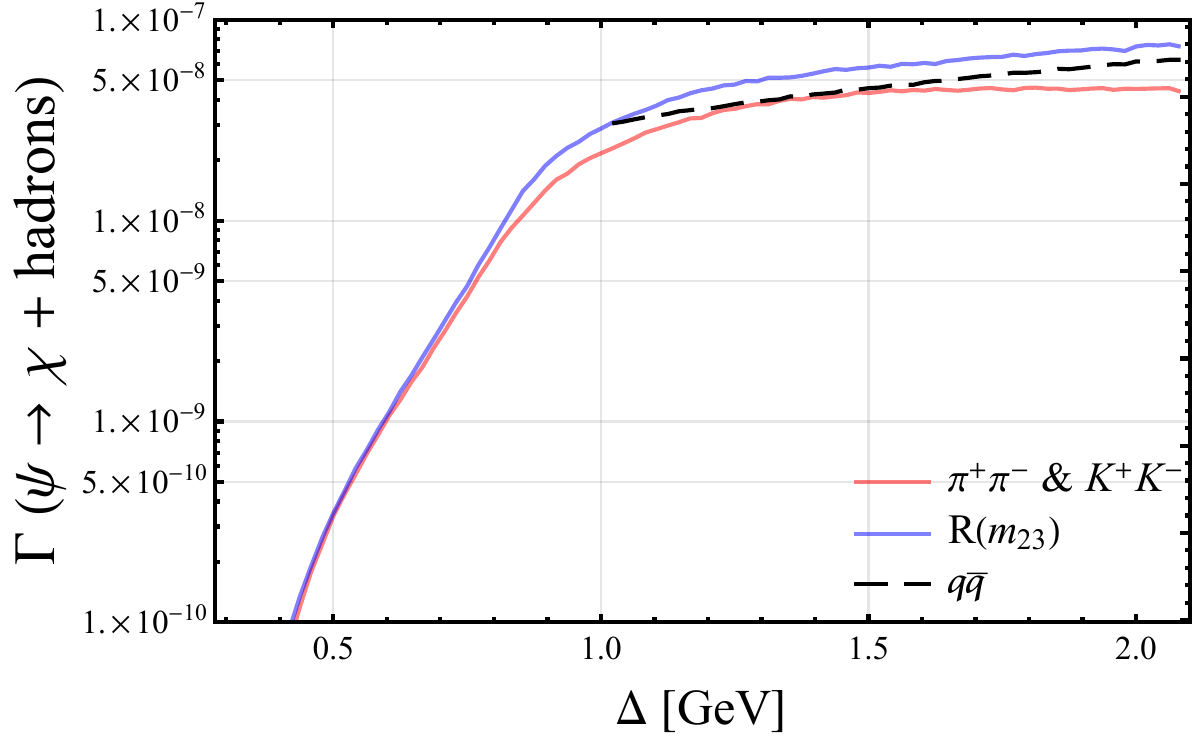}
    \caption{Comparison of the hadronic decay width of $\psi$ estimated using three-body decays $\chi\mathfrak{m}^+\mathfrak{m}^-$ (red), $R(s)$ data (blue) and the parton level calculation (dashed black). Common factors of $\epsilon^2 e^2 g^2_D m^2/\Delta^2$ have been omitted.}
    \label{fig: HC_compare_hadronic}
\end{figure}

In the case of the hypercharge portal, hadronic decays are relevant when $\Delta \equiv m_\psi - m_\chi \lesssim 1~\mathrm{GeV}$. At the lowest order\footnote{Two-body decays such as $\psi \rightarrow \chi \mathfrak{m}^{0}$ proceed through an off-shell $Z$. The corresponding decay widths are suppressed by a factor of $m^4_{Z'}/m^4_{Z}$ and can be neglected}, we have three-body decays of the form $\psi \rightarrow \chi \mathfrak{m}^+\mathfrak{m}^-$. The squared matrix element for these decay channels is given by
\begin{equation}
    \begin{split}
        \overline{|\mathcal{M}|^2} = \frac{\epsilon^2 e^2 g^2_D m^2 |F_\mathfrak{m}(m^2_{23})|^2}{\Delta^2 (m^2_{23}-m^2_{Z'})^2}&\left[(m^2_{12}-m^2_{13})^2 -(m^2_{23}-4m^2_\mathfrak{m})(\Delta^2 - m^2_{23})\right] \;.
    \end{split}
\end{equation}
For $\Delta < 2~\mathrm{GeV}$, only $\mathfrak{m} = \pi, K$ are kinematically accessible. We compare our results against an estimate based on the ratio $R(s) = \sigma(e^+ e^- \rightarrow \mathrm{hadrons}; s)/\sigma(e^+ e^- \rightarrow \mu^+ \mu^-; s)$
\begin{equation}
    d\Gamma (\psi \rightarrow \chi + \mathrm{hadrons}) \approx d\Gamma(\psi \rightarrow \chi \mu^+ \mu^-)\times R(m_{23}) \;.
\label{eq: gamma_approx_R_psi}
\end{equation}
We note that the correlations between the leptonic and hadronic matrix elements in the decays of $\psi$ are not the same as in the collision of $e^+ e^-$, due to the masses $m_\psi, m_\chi$. More precisely, in the case of $e^+ e^-$ collisions, we have
\begin{equation}
    d\sigma(e^+ e^- \rightarrow \mathrm{hadrons}) \propto \frac{E^{\mu\nu}H_{\mu\nu}}{Q^6} d\Phi_\mathrm{had}(Q) \;,
\end{equation}
where $Q$ is the invariant mass of the hadronic final state, $E^{\mu\nu} = -g^{\mu\nu}\frac{Q^2}{2} + k^\mu_1k^\nu_2 + k^\nu_1k^\mu_2$ and $H_{\mu\nu}$ is the hadronic tensor containing the relevant form factors and momenta of the hadronic final state. In the case of $\psi$ decays, we have
\begin{equation}
    d\Gamma \propto \frac{X^{\mu\nu}H_{\mu\nu}}{m_\psi (Q^2 - m^2_{Z'})^4} d\Phi_2(m_\psi; m_\chi, Q)dQ^2d\Phi_\mathrm{had}(Q) \;,
\end{equation}
where $X^{\mu\nu} = -g^{\mu\nu}(\frac{\Delta^2 - Q^2}{2}) + p^\mu_0 p^\nu_1+ p^\nu_0 p^\mu_1$. Since in general, $H_{\mu\nu}$ is dependent on the momenta of the hadrons in the final state, Eq.~\ref{eq: gamma_approx_R_psi} does not accurately capture the correlations between the two tensors. As shown in Fig.~\ref{fig: HC_compare_hadronic}, the parton-level calculation lies within the range predicted by these two approaches for $\Delta \gtrsim 1~\mathrm{GeV}$. The $\mathcal{O}(1)$ difference between the two approaches is likely due to the presence of the $\pi^+\pi^-\pi^0$ channel, which is resonantly produced at $Q = m_\omega \sim 0.75~\mathrm{GeV}$ and $Q = m_\phi \sim 1~\mathrm{GeV}$. We transition from the hadronic calculation to the parton-level calculation at the value of $\Delta$ where the two calculations match.

\section{Photon Yields and DOM Efficiency}
\label{sec: DOM_Efficiency}

\begin{figure}[h!]
    \centering
    \includegraphics[width=0.5\linewidth]{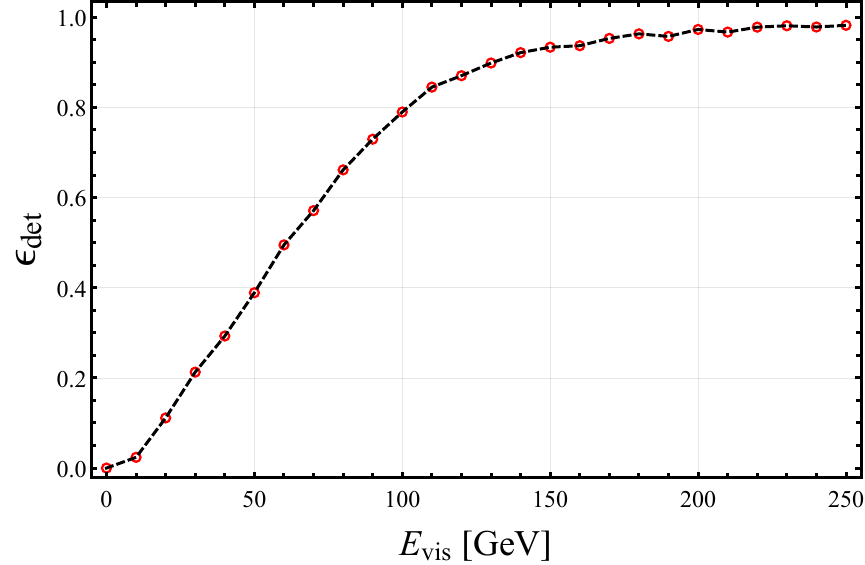}
    \caption{Detection efficiency $\epsilon_\mathrm{det}$ as a function of the deposited energy $E_\mathrm{vis}$ for electromagnetic and hadronic cascades, assuming a point-source emission of Cherenkov photons.}
    \label{fig: det_eff}
\end{figure}

For both of the models considered in this paper, we focus exclusively on the detection of electromagnetic and hadronic cascades, as opposed to muon tracks. In the case of cascade topologies, the profile of Cherenkov photons can be modeled as a point-source emission~\cite{IceCube:2013dkx}. The total number of photons emitted by the cascade scales linearly with respect to the energy deposited in the cascade,
\begin{equation}
    n_0 \approx 1.7 \times 10^{5} \left(\frac{E_\mathrm{vis}}{\mathrm{GeV}}\right) \;.
\end{equation}
As shown in Equation~5 of~\cite{IceCube:2013dkx}, the expected number of photons at a distance $r$ from the cascade can be estimated as
\begin{equation}
    n(r) = \frac{n_0 A_\mathrm{eff}}{4\pi r \lambda_c} \times\frac{e^{-r/\lambda_p}}{\tanh\left({\frac{r}{\lambda_c}}\right)} \;,
\end{equation}
where $\lambda_p = 26~\mathrm{m}$, $\lambda_c = 10~\mathrm{m}$ and $A_\mathrm{eff} = 80~\mathrm{cm^2}$ is the effective photon collection area of the photomultiplier tubes (PMTs) used in the DOMs~\cite{IceCube:2010dpc}. Given the expected number of photons $n$, the actual number of photons that arrive at the DOM follows a Poisson distribution, i.e., $P(k) = \frac{n^k}{k!} e^{-n}$, which we use for sampling. Additionally, since the photon collection area of the PMT only covers the bottom hemisphere of the DOM, we only consider DOMs located at depths above the cascade. We perform a Monte Carlo analysis to estimate the detection efficiency of cascades as a function of the deposited energy $E_\mathrm{vis}$. We use the 86-string configuration of IceCube (IC86), which includes the 6 HQE DeepCore strings. For a given value of $E_\mathrm{vis}$, we uniformly sample $N_\mathrm{sim}=10^5$ number of locations for the cascade inside the detector volume. Then, we consider a cascade to be detected if it results in at least one photon in four different DOMs, corresponding to the SMT4 trigger. The results are shown in Fig.~\ref{fig: det_eff}, which we use in our analysis.

% Bibliography

%% [A] Recommended: using JHEP.bst file
%% \bibliographystyle{JHEP}
%% \bibliography{biblio.bib}

%% or
%% [B] Manual formatting (see below)
%% (i) We suggest to always provide author, title and journal data or doi:
%% in short all the informations that clearly identify a document.
%% (ii) please avoid comments such as "For a review'', "For some examples",
%% "and references therein" or move them in the text. In general, please leave only references in the bibliography and move all
%% accessory text in footnotes.
%% (iii) Also, please have only one work for each \bibitem.

\bibliographystyle{JHEP}
\bibliography{biblio}
\end{document}